\documentclass
[superscriptaddress,secnumarabic,amssymb,amsmath,nobibnotes,aps,prd,showkeys,showpacs,nofootinbib,onecolumn]{revtex4}%
\usepackage{lscape}
\usepackage{graphicx}
\usepackage{epsf}
\usepackage{bm}
\usepackage{amsmath}
\usepackage{amsfonts}
\usepackage{amssymb}
\usepackage{epstopdf}
\usepackage{mwe}
\usepackage{caption}
\usepackage{subfigure}%
\providecommand{\U}[1]{\protect\rule{.1in}{.1in}}

\newcommand{\be}{\begin{equation}}
\newcommand{\ee}{\end{equation}}

\newcommand{\mincir}{\raise
-3.truept\hbox{\rlap{\hbox{$\sim$}}\raise4.truept\hbox{$<$}\ }}
\newcommand{\magcir}{\raise
-3.truept\hbox{\rlap{\hbox{$\sim$}}\raise4.truept\hbox{$>$}\ }}

\ifx\pdfoutput\relax\let\pdfoutput=\undefined\fi
\newcount\msipdfoutput
\ifx\pdfoutput\undefined\else
\ifcase\pdfoutput\else
\ifx\paperwidth\undefined\else
\ifdim\paperheight=0pt\relax\else\pdfpageheight\paperheight\fi
\ifdim\paperwidth=0pt\relax\else\pdfpagewidth\paperwidth\fi
\fi\fi\fi
\begin{document}
\title{Dynamical Analysis of an Integrable Cubic Galileon Cosmological Model}
\author{Alex Giacomini}
\email{alexgiacomini@uach.cl}
\affiliation{Instituto de Ciencias F\'{\i}sicas y Matem\'{a}ticas, Universidad Austral de
Chile, Valdivia, Chile}
\author{Sameerah Jamal}
\email{sameerah.jamal@wits.ac.za}
\affiliation{School of Mathematics and Centre for Differential Equations,Continuum
Mechanics and Applications, University of the Witwatersrand, Johannesburg,
South Africa}
\author{Genly Leon}
\email{genly.leon@pucv.cl}
\affiliation{Instituto de F\'{\i}sica, Pontificia Universidad Cat\'olica de
Valpara\'{\i}so, Casilla 4950, Valpara\'{\i}so, Chile}
\author{Andronikos Paliathanasis}
\email{anpaliat@phys.uoa.gr}
\affiliation{Instituto de Ciencias F\'{\i}sicas y Matem\'{a}ticas, Universidad Austral de
Chile, Valdivia, Chile}
\affiliation{Institute of Systems Science, Durban University of Technology, PO Box 1334,
Durban 4000, Republic of South Africa}
\author{Joel Saavedra}
\email{joel.saavedra@pucv.cl}
\affiliation{Instituto de F\'{\i}sica, Pontificia Universidad Cat\'olica de
Valpara\'{\i}so, Casilla 4950, Valpara\'{\i}so, Chile}

\begin{abstract}
Recently a cubic Galileon cosmological model was derived by the assumption
that the field equations are invariant under the action of point
transformations. The cubic Galileon model admits a second conservation law
which means that the field equations form an integrable system. The analysis
of the critical points for this integrable model is the main subject of this
work. To perform the analysis, we work on dimensionless variables different
from that of the Hubble normalization. New critical points are derived while
the gravitational effects which follow from the cubic term are studied.

\end{abstract}
\maketitle
\date{\today}

\section{Introduction}

A theory which has drawn the attention of the scientific society in the last
few years is the Galileon gravity \cite{nik,gal02}. It belongs to the modified
theories of gravity in which a noncanonical scalar field is introduced and the
field equations are invariant under the Galilean transformation. The Action
integral of the Galileon gravity belongs to the Horndeski theories \cite{hor},
which means that the gravitational theory is of second-order \cite{m10}. The
vast applications of study for Galileons in the literature covers all the
areas of gravitation physics from neutron stars, black holes, acceleration of
the universe, for instance see
\cite{cha,char,chr2,chr3,adolfo1,adolfo2,bar,bellini,bartolo,babi,Bhat,genlyGL,genlyGL2,m07,m08,m09,m10,m11}
and references therein.

In this work we are interested in the cosmological scenario and specifically
in the so-called Galileon cosmology \cite{chow,defelice,cedric}. In cosmology,
the Galileon field has been applied in order to explain various phases of the
evolution of the universe \cite{Teg,Kowal,Komatsu,Ade15,planck2015}.
Specifically the new terms in the gravitational Action integral can force the
dynamics in a way such that the model fits the observations. The mechanics can
explain the inflation era \cite{inf1,inf2,inf3,inf4,inf5,inf6,inf7} as also
the late time-acceleration of the universe
\cite{late1,late2,late3,late4,late5,late6}. Last but not least the growth
index of matter perturbations have been constrained in \cite{matter1}.

As we mentioned in the previous paragraph, Galileon gravity belongs to the
Hondenski theories and specifically there is an infinite number of different
models which can be constructed from a general Lagrangian. A simple model is
the cubic Galileon model \cite{bar,bellini,bartolo,babi,Bhat,genlyGL,genlyGL2}
where the\ Action Integral is that of a canonical scalar field plus a new term
which has a cubic derivative on the Galileon field. The theory can be seen as
a first extension of the scalar field cosmology. Due to this cubic term, the
nonlinearity and the complexity of the field equations is increased
dramatically. Recently in \cite{Dimakis:2017kwx} a cubic Galileon model was
derived which admits an additional conservation law and where the field
equations formed an integrable dynamical system.

Integrability is an important issue in all the areas of physics and
mathematical sciences. The reason is that while a dynamical system can be
studied numerically, it is unknown if an actual solution which describes the
\textquotedblleft orbits\textquotedblright\ exist. The integrable cubic
Galileon model admits special solutions which describes an ideal gas universe,
that is, power-law scale factors. While this is similar to the solution of the
canonical scalar field, we found that the power of the power-law solution is
not strongly constrained by the Galileon field and that is because of the
cubic term. On the other hand a special property of that model is that when
the potential in the Action integral dominates, then the cubic term disappears
which mean that the theory approach {is} that of a canonical scalar field.
However as we shall see from our analysis, the existence of the conservation
law provides new dynamics which have not been investigated previously.

The scope of this work is to study the evolution of the integrable cubic
Galileon cosmological model. For that we perform an analysis of the critical
points. In particular, in Section \ref{sec:level1} we briefly discuss the
cubic Galileon cosmology and we review the integrable case that was derived
before in \cite{Dimakis:2017kwx}. Section \ref{dynaev} includes the main
material of our analysis where we rewrite the field equations in dimensionless
variables. We define variables different from that of the $H$-normalizaton
where we find that the dynamical system is not bounded. Because of the latter
property the critical points at the finite and the infinite regions are
studied. At the finite region we find various critical points which can
describe the expansion history of our universe as also the matter dominated
era. Appendices \ref{analysisQ1}, \ref{appendixA}, \ref{AppP2}, \ref{AppP4},
\ref{AppP14} and \ref{AppP15-P16} include important mathematical material
which justify our analysis. One important property of the integrable model
that we study is that there is a limit in which the terms in the field
equations (which follow from the cubic term of the Galileon Lagrangian) vanish
and the model is then reduced to that of a canonical scalar field cosmological
model. Hence, in order to study the effects of the cubic term in Section
\ref{asym} we perform an asymptotic expansion of the solution when the cubic
term dominates the universe. In Section \ref{matter} we extend our analysis to
the case where an extra matter term is included in the gravitational Action
integral. Finally, we discuss our results and draw our conclusions in Section
\ref{conc}.

\section{Cubic Galileon Cosmology}

\label{sec:level1}

The cubic Galileon model is defined by the following Action integral
\begin{equation}
S=\int d^{4}x\sqrt{-g}\left(  \frac{1}{2}R-\frac{1}{2}\partial^{\mu}%
\phi\partial_{\mu}\phi-V(\phi)-\frac{1}{2}g(\phi)\partial^{\mu}\phi
\partial_{\mu}\phi\Box\phi\right)  \label{lan.03}%
\end{equation}
which has various cosmological and gravitational applications.

In the cosmological scenario of a homogeneous and isotropic universe with zero
spatially curvature the line element of the spacetime is that of the FLRW
metric%
\begin{equation}
ds^{2}=-dt^{2}+a^{2}\left(  t\right)  \left(  dx^{2}+dy^{2}+dz^{2}\right)  ,
\label{bd.05}%
\end{equation}
where $a\left(  t\right)  $ is the scale factor of the universe.

Indeed for this line element, variation with respect to the metric tensor in
(\ref{lan.03}) provides the gravitational field equations
\begin{equation}
3H^{2}=\frac{\dot{\phi}^{2}}{2}\left(  \,1-6g(\phi)H\dot{\phi}+g^{\prime}%
(\phi)\,\dot{\phi}^{2}\right)  +V(\phi), \label{ss1}%
\end{equation}
and
\begin{equation}
2\dot{H}+\dot{\phi}^{2}\left(  1+g^{\prime}(\phi)\,\dot{\phi}^{2}%
-3g(\phi)\,H\,\dot{\phi}+g(\phi)\,\ddot{\phi}\right)  =0, \label{f1}%
\end{equation}
while variation with respect to the field $\phi$, provides the (modified)
\textquotedblleft Klein-Gordon\textquotedblright\ equation
\begin{equation}
\ddot{\phi}\left(  2\dot{\phi}^{2}\,g^{\prime}(\phi)-6H\,g(\phi)\,\dot{\phi
}+1\right)  +\dot{\phi}^{2}\left(  \frac{1}{2}\dot{\phi}^{2}\,g^{\prime\prime
}(\phi)-3\,g(\phi)\,\dot{H}-9\,H^{2}\,g(\phi)\right)  +3\,H\,\dot{\phi
}+V^{\prime}(\phi)=0, \label{f2}%
\end{equation}
which describes the evolution of the field, and $H=\frac{\dot{a}}{a}$. Recall
that we have assumed that the Galileon field inherits the symmetries of the
spacetime, that is if $K^{\mu}~$is an isometry of (\ref{bd.05}), i.e. $\left[
K,g_{\mu\nu}\right]  =0$, then $\phi$ inherits the symmetries of the spacetime
if and only if $\left[  K,\phi\right]  =0$. Therefore $\phi$ is only a
function of the \textquotedblleft$t$\textquotedblright\ parameter, i.e.
$\phi\left(  x^{\mu}\right)  =\phi\left(  t\right)  $.

An equivalent way to write the field equations is by defining fluid components
such as energy density and pressure which corresponds to the Galileon field.
Indeed if we consider the energy density
\begin{equation}
\rho_{G}=\frac{\dot{\phi}^{2}}{2}\left(  \,1-6g(\phi)H\dot{\phi}+g^{\prime
}(\phi)\,\dot{\phi}^{2}\right)  +V(\phi),
\end{equation}
and the pressure term
\begin{equation}
p_{G}=\frac{\dot{\phi}^{2}}{2}\left(  1+2g\ddot{\phi}+g_{,\phi}\dot{\phi}%
^{2}\right)  -V(\phi),
\end{equation}
the field equations take the form $G_{~\nu}^{\mu}=T_{~\nu}^{\left(  G\right)
\mu}$, where $T_{\mu\nu}^{\left(  G\right)  }$ is the energy momentum tensor
corresponding to the Galileon field and
\begin{equation}
T_{\mu\nu}^{\left(  G\right)  }=\rho_{G}u_{\mu}u_{\nu}+p_{G}\left(  g_{\mu\nu
}+u_{\mu}u_{\nu}\right)
\end{equation}
in which $u^{\mu}=\delta_{t}^{\mu}$ is the normalized comoving observer
$\left(  u^{\mu}u_{\mu}=-1\right)  $. Last but not least equation (\ref{f2})
is now equivalent to the Bianchi identity $T_{~~~\ ~~~\ \ ;\nu}^{\left(
G\right)  \mu\nu}=0$, that is,%
\begin{equation}
\dot{\rho}_{DE}+3H(\rho_{DE}+p_{DE})=0.
\end{equation}

Last but not least the dark energy equation-of-state parameter is defined as
follows.
\begin{equation}
w_{DE}\equiv\frac{p_{DE}}{\rho_{DE}}=\frac{1}{3H^{2}}\left[  \frac{\dot{\phi
}^{2}}{2}\left(  1+2g\ddot{\phi}+g_{,\phi}\dot{\phi}^{2}\right)
-V(\phi)\right]  . \label{EoS}%
\end{equation}
It can be shown that with a proper election of $g(\phi)$, the
equation-of-state parameter $w_{DE}$ can realize the quintessence scenario,
the phantom one and cross the phantom divide during the evolution, which is
one of the advantages of Galileon cosmology. In general the specific functions
of $g\left(  \phi\right)  ~$and $V\left(  \phi\right)  $ are unknown and for
different functions there will be a different evolution.

\subsection{The extra conservation law}

Recently in \cite{Dimakis:2017kwx} two unknown functions were specified by the
requirement that the gravitational field equations form an integrable
dynamical system. Specifically the following functions were found
\begin{equation}
V(\phi)=V_{0}e^{-\lambda\phi}\quad\text{and}\quad g(\phi)=g_{0}e^{\lambda\phi
}. \label{pot}%
\end{equation}
There exists a symmetry vector which provides, with the use of Noether's
second theorem, the following conservation law for the field equations
\begin{equation}
I_{1}=-\left(  {2\,a^{2}\dot{a}{{-}}\frac{{{2}}}{\lambda}a^{3}\dot{\phi}%
}\right)  {+g_{0}{\mathrm{e}^{\lambda\phi}}a^{3}\dot{\phi}^{3}-}\frac
{6}{\lambda}g_{0}a^{2}e^{\lambda\phi}\dot{a}\dot{\phi}^{2}. \label{con01}%
\end{equation}

Because of the nonlinearity of the field equations the general solution cannot
be written in a closed-form. However from the symmetry vector invariant curves
have been defined and by using the zero-th order invariants some power law
(singular solutions) have been derived. In particular the following solutions
were obtained:
\begin{equation}
a_{1}\left(  t\right)  =a_{0}t^{p}~,~\phi\left(  t\right)  =\frac{2}{\lambda
}\ln\left(  \phi_{0}t\right)  ~,~g_{0}=\frac{\lambda\left(  2-\lambda
^{2}p\right)  }{4(3p-1)\phi_{0}^{2}},~V_{0}=\phi_{0}^{2}\left(  \frac
{2}{\lambda^{2}}+p(3p-2)\right)  \label{solution1}%
\end{equation}
and
\begin{equation}
a_{2,3}\left(  t\right)  =a_{0}t^{\frac{1}{3}}~,~\phi_{2,3}\left(  t\right)
=\pm\frac{\sqrt{6}}{3}\ln(\phi_{0}t)~,~V_{0}=0~,~\lambda_{2,3}=\pm\sqrt{6}.
\end{equation}
These solutions are special solutions since they exist for specific initial
conditions. In order to study the general evolution of the system we perform
an analysis in the phase-space.

A phase-space analysis for this cosmological model has been performed
previously in \cite{genlyGL}, however the integrable case with $g\left(
\phi\right)  $ and $V\left(  \phi\right)  $ given by the expressions
(\ref{pot}) were excluded from \cite{genlyGL}. Moreover there is a special
observation in the integrable case in the sense that $V\left(  \phi\right)
g\left(  \phi\right)  =const$. The latter means that when $V\left(
\phi\right)  $ dominates, $g\left(  \phi\right)  $ becomes very small and the
cubic Galileon model reduces to that of a canonical scalar field which can
mimic also the cosmological constant when $V\left(  \phi\right)  \gg \dot{\phi
}^{2}$.

In the following we write the field equations in new dimensionless variables
and we perform our analysis.

\section{Evolution of the dynamical system}

\label{dynaev}

From equation \eqref{ss1} one immediately sees that the Hubble function
$H\left(  t\right)  $ can cross the value $H\left(  t\right)  =0$, from
negative to positive values, or vice-verca. This means that the standard
$H$-normalizaton is not useful and new variables have to be defined. We
introduce the new variables
\begin{equation}
x=\frac{\dot{\phi}}{\sqrt{6(H^{2}+1)}},y=\frac{V_{0}e^{-\lambda\phi}}%
{3(H^{2}+1)},z=\frac{H}{\sqrt{H^{2}+1}}, \label{eq.37}%
\end{equation}
and the parameter $\alpha=g_{0}V_{0}$ so we obtain the three-dimensional
dynamical system%

\begin{equation}
x^{\prime}=\frac{f_{1}(\mathbf{x})}{f_{4}(\mathbf{x})},~y^{\prime}=\frac
{f_{2}(\mathbf{x})}{f_{4}(\mathbf{x})},~z^{\prime}=\frac{f_{3}(\mathbf{x}%
)}{f_{4}(\mathbf{x)}} \label{eq.38}%
\end{equation}
where~$\mathbf{x=}\left(  x,y,z\right)  $ and functions $f_{1}-f_{4}$
are\thinspace\ defined by the following expressions%

\begin{subequations}
\label{eq.38b}%
\begin{align}
&  f_{1}(x,y,z)=y^{2}\left(  3x^{3}z+3xz\left(  -y+z^{2}-2\right)  +\sqrt
{6}\lambda y\right)  +6\alpha^{2}x^{5}\left(  2\lambda^{2}x^{2}z-\sqrt
{6}\lambda x-2\sqrt{6}\lambda xz^{2}+6z^{3}\right) \nonumber\\
&  +\alpha x^{2}y\left(  18\lambda x^{3}z-\sqrt{6}x^{2}\left(  2\lambda
^{2}+12z^{2}+3\right)  -6\lambda xz\left(  y-2z^{2}\right)  +3\sqrt{6}\left(
2yz^{2}+y-2z^{4}+z^{2}\right)  \right)  ,\\
&  f_{2}(x,y,z)=2y\left(  12\alpha^{2}\lambda^{2}x^{6}z-6\sqrt{6}\alpha
^{2}\lambda x^{5}\left(  2z^{2}+1\right)  +18\alpha x^{4}z\left(  \lambda
y+2\alpha z^{2}\right)  -4\sqrt{6}\alpha x^{3}y\left(  \lambda^{2}%
+3z^{2}\right)  \right. \nonumber\\
&  \left.  +3x^{2}yz\left(  y-2\alpha\lambda\left(  y-2\left(  z^{2}+1\right)
\right)  \right)  +\sqrt{6}xy\left(  6\alpha z^{2}\left(  y-z^{2}\right)
-\lambda y\right)  +3y^{2}z\left(  z^{2}-y\right)  \right)  ,\\
&  f_{3}(x,y,z)=3\left(  z^{2}-1\right)  \left(  y^{2}\left(  x^{2}+2\alpha
x\left(  \sqrt{6}z-\lambda x\right)  +z^{2}\right)  +4\alpha^{2}x^{4}\left(
\lambda^{2}x^{2}-\sqrt{6}\lambda xz+3z^{2}\right)  \right. \nonumber\\
&  \left.  +2\alpha xy\left(  3\lambda x^{3}-2\sqrt{6}x^{2}z+2\lambda
xz^{2}-\sqrt{6}z^{3}\right)  -y^{3}\right)  ,\\
&  f_{4}(x,y,z)=2\left(  2\alpha x\left(  3\alpha x^{3}+2\lambda xy-\sqrt
{6}yz\right)  +y^{2}\right)  ,
\end{align}
and prime denotes the new derivative $\frac{df}{d\tau}\equiv f^{\prime}%
\equiv\frac{\dot{f}}{\sqrt{H^{2}+1}}$. \newline Interestingly, the system
\eqref{eq.38} admits the first integral
\end{subequations}
\begin{equation}
2\alpha x^{3}\left(  \lambda x-\sqrt{6}z\right)  +y(x^{2}-z^{2})+y^{2}=0
\label{restr_1}%
\end{equation}
which is the Friedmann's first equation and constrains the evolution of the
solution. Thus, the dynamics are restricted to a surface given by
\eqref{restr_1}. For a fixed value of $y$, the first and last equations in
\eqref{eq.38} are invariant under the discrete symmetry $(x,z,\tau
)\rightarrow-(x,z,\tau)$. Thus, the fixed points related by this discrete
symmetry have the opposite dynamical behavior. By definition, $y\geq0$.

Let's compare with the variables introduced in \cite{genlyGL} defined by
$x_{1}=\frac{\dot{\phi}}{\sqrt{6}H},\quad y_{1}=\frac{\sqrt{V(\phi)}}{\sqrt
{3}H},\quad z_{1}=g(\phi)H\dot{\phi}.$

Since we have chosen here $V(\phi)$ and $g(\phi)$ such that $g(\phi
)V(\phi)=\alpha$ we have the relations $x_{1}=\frac{x}{z},\quad y_{1}%
=\frac{\sqrt{y}}{z},\quad z_{1}=\frac{\sqrt{\frac{2}{3}}\alpha xz}{y},$ and
the extra relation $\alpha x_{1}-\sqrt{\frac{3}{2}}y_{1}^{2}z_{1}=0.$ This
implies that the fixed points $A^{\pm},B^{\pm},C$ and $D$ investigated in
detail in \cite{genlyGL}, do not exist in our scenario since the values of
their coordinates $(x_{1},y_{1},z_{1})$ do not satisfy the above extra relation.

Furthermore, some cosmological parameters with great physical significance are
the effective equation of state parameter $w_{tot}\equiv\frac{p_{tot}}%
{\rho_{tot}}=w_{DE}$ (because we set $\rho_{m}=0$) and the `deceleration
parameter\textquotedblright\
\begin{equation}
q\equiv-1-\frac{\dot{H}}{H^{2}}=\frac{1}{2}+\frac{3}{2}w_{tot}. \label{decc0}%
\end{equation}
Some conditions for the cosmological viability of the most general
scalar-tensor theories has to be satisfied by extended Galileon dark energy
models; say the model must be free of ghosts and Laplacian instabilities
\cite{DeFelice:2010pv,DeFelice:2011bh,Appleby:2011aa}. In the special case of
the action \eqref{lan.03} (in units where $\kappa\equiv8\pi G=1$), we require
for the avoidance of Laplacian instabilities associated with the scalar field
propagation speed that \cite{DeFelice:2011bh}
\begin{align}
&  c_{S}^{2}\equiv\frac{6w_{1}H-3w_{1}^{2}-6\dot{w}_{1}}{4w_{2}+9w_{1}^{2}%
}\geq0;\\
&  w_{1}\equiv g\dot{\phi}^{3}+2{H},w_{2}\equiv3\dot{\phi}^{2}\left[  \frac
{1}{2}+g_{,\phi}\dot{\phi}^{2}-6Hg\dot{\phi}\right]  -{9H^{2}}. \label{cscon}%
\end{align}
Meanwhile, for the absence of ghosts it is required that
\begin{equation}
Q_{S}\equiv\frac{(4w_{2}+9w_{1}^{2})}{3w_{1}^{2}}>0. \label{Qscon}%
\end{equation}
Finally, we have from the Eqs. \eqref{EoS}, \eqref{cscon} and \eqref{Qscon}
that the phantom phase can be free of instabilities and thus cosmologically
viable, as it was already shown for Galileon cosmology \cite{DeFelice:2011bh}.

\subsection{Analysis at the finite region}

The fixed points/ fixed lines at the finite region of the system
\eqref{eq.38}, and a summary of their stability conditions are presented in
table \ref{tab:Table_I}. In table \ref{tab:Table_II} we display several
cosmological parameters for the fixed points at the finite region of the
system \eqref{eq.38}. The discussion about the physical interpretation of
these points and the points at infinity is left for section
\ref{discussion-points} \newline

\begin{table}[th]
\caption{Summary of the stability conditions of the fixed points at the finite
region of the system \eqref{eq.38}. Where $\mathcal{P}(x)=6(1+\alpha\lambda)
{x}^{3}+3\sqrt{6}(2\alpha+\lambda){x}^{2} +2(3+\lambda^{2})x +\sqrt{6}\lambda
$. We used the acronyms ``A.S.'' for Asymptotically Stable and ``A.U.'' for
Asymptotically Unstable.}%
\label{tab:Table_I}
\centering \resizebox{\textwidth}{!}{
\begin{tabular}
[c]{cccc}\hline\hline
\text{Label}: \text{Coordinates} $(x,y,z)$ & \text{Existence} &
\text{Eigenvalues} & \text{Stability}\\\hline
$P_{1}:\left(  0,0,0\right)  $ & \text{Always} & \text{Undetermined} &
\text{Numerical analysis}\\
$P_{2}:\left(  -\frac{\sqrt{6}}{\lambda},0,-1\right)  $ & \text{Always} &
$0,-3,-6$ & stable for
$\lambda<-\sqrt{6}$ \\
& & & or $0<\lambda<\sqrt{6}$, \\
& & & saddle for $-\sqrt{6}<\lambda<0$ \\
& & & or $\lambda>\sqrt{6}$.\\
& & & (see appendix \ref{AppP2})\\
$P_{3}:\left(  \frac{\sqrt{6}}{\lambda},0,1\right)  $ & \text{Always} &
$0,3,6$ & unstable for
$\lambda<-\sqrt{6}$ \\
& & & or $0<\lambda<\sqrt{6}$, \\
& & & saddle for $-\sqrt{6}<\lambda<0$ \\
& & & or $\lambda>\sqrt{6}$.\\
& & & (see appendix \ref{AppP2})\\
$P_{4}:\left(  0, 1 ,1\right)  $ & $\lambda=0$ & $-3, -3, 0$ &
\text{A.S.} (see Appendix \ref{AppP4})\\
$P_{5}:\left(  0,1,-1\right)  $ & $\lambda=0$ & $3, 3, 0$ &
\text{A.U.} (see Appendix \ref{AppP4})\\
$P_{6}(x_{c}):\left(  x_{c},\sqrt{\frac{2}{3}}\lambda x_{c}+x_{c}^{2}+1,-1\right)  $ & Where $x_{c}\neq0$ is a real root of $\mathcal{P}(x)$ &
& \\
& such that $\sqrt{\frac{2}{3}}\lambda x_{c}+x_{c}^{2}+1\geq0$ &
\text{Numerical analysis} & \text{Numerical analysis}\\
$P_{7}(x_{c}):\left(  -x_{c},\sqrt{\frac{2}{3}}\lambda x_{c}+x_{c}^{2}+1,1\right)  $ & Where $x_{c}\neq0$ is a real root of $\mathcal{P}(x)$ &
& \\
& such that $\sqrt{\frac{2}{3}}\lambda x_{c}+x_{c}^{2}+1\geq0$ &
\text{Numerical analysis} & \text{Numerical analysis}\\
$P_{8}:\left(  0,0,-1\right)  $ & \text{Always} & $-3,-3, -\left(  6
\alpha\lambda-\frac{3}{2}\right)  $ & \text{Sink for} $\alpha\lambda>\frac
{1}{4}$\\
&  &  & \text{Saddle otherwise}\\
$P_{9}:\left(  0,0,1\right)  $ & \text{Always} & $3,3, \left(  6 \alpha
\lambda-\frac{3}{2}\right)  $ & \text{Source} for $\alpha\lambda>\frac{1}{4}$\\
&  &  & \text{Saddle otherwise}\\
$P_{10}: \left(  -\frac{\sqrt{\frac{3}{2}}}{\lambda}, 0 , -1\right)  $ &
$\alpha=0, \lambda\neq0$ & $\left(  -\frac{9}{2 \lambda^{2}},\frac{3}{4}
\left(  2-\frac{9}{\lambda^{2}}\right)  ,-\frac{9}{2 \lambda^{2}}-3\right)  $
& sink for $0<\lambda^{2}<\frac{3}{2}$\\
&  &  & saddle otherwise\\
$P_{11}: \left(  \frac{\sqrt{\frac{3}{2}}}{\lambda}, 0, 1\right)  $ &
$\alpha=0, \lambda\neq0$ & $\left(  \frac{9}{2 \lambda^{2}},-\frac{3}{4}
\left(  2-\frac{9}{\lambda^{2}}\right)  ,\frac{9}{2 \lambda^{2}}+3\right)  $ &
source for $0<\lambda^{2}<\frac{3}{2}$\\
&  &  & saddle otherwise\\
$P_{12}: \left(  -\frac{\sqrt{\frac{3}{2}}}{\lambda}, \frac{3}{2\lambda^{2}},
-1\right)  $ & $\lambda^{2}+3\alpha\lambda-3=0$ & $0,3,-\frac{3}{2}$ &
saddle\\
$P_{13}: \left(  \frac{\sqrt{\frac{3}{2}}}{\lambda}, \frac{3}{2\lambda^{2}},
1\right)  $ & $\lambda^{2}+3\alpha\lambda-3=0$ & $0,3,-\frac{3}{2}$ & saddle\\
$P_{14}(z_{c}):\left(  0, z_{c}^{2}, z_{c}\right)  $ & $\lambda=0$ & $\left(
0,-3 z_{c},-3 z_{c}\right)  $ & A.S. for $0<{z_c}\leq 1$\\
& & & A.U. for $-1\leq {z_c}<0$ \\
& & & (see Appendix \ref{AppP14}).\\
$P_{15}(z_{c}):\left(\beta{z_{c}},\sqrt{6}\alpha\beta{z_{c}}^2,z_{c}\right)   $ &
$\lambda=0, \beta= \frac{1}{\sqrt{2}} \left(  \sqrt{3}\alpha-\sqrt
{3\alpha^{2}-2}\right), \alpha >\sqrt{\frac{2}{3}}$& $\left(  0,-3 z_{c},-3 z_{c}\right)  $ & A.S. for $0<{z_c}\leq 1$\\
& & & A.U. for $-1\leq {z_c}<0$ \\
& & & (see Appendix \ref{AppP15-P16}).\\
$P_{16}(z_{c}):\left(\beta{z_{c}},\sqrt{6}\alpha\beta{z_{c}}^2,z_{c}\right)    $ &
$\lambda=0, \beta= \frac{1}{\sqrt{2}} \left(  \sqrt{3}\alpha+\sqrt
{3\alpha^{2}-2}\right),  \alpha >\sqrt{\frac{2}{3}}$ & $\left(  0,-3 z_{c},-3 z_{c}\right)  $ & A.S. for $0<{z_c}\leq 1$\\
& & & A.U. for $-1\leq {z_c}<0$ \\
& & & (see Appendix \ref{AppP15-P16}).\\\\\hline\hline
\end{tabular}
}\end{table}

\begin{table}[th]
\caption{Cosmological parameters for the fixed points at the finite region of
the system \eqref{eq.38}. }%
\label{tab:Table_II}
\centering \resizebox{\textwidth}{!}{
\begin{tabular}
[c]{ccc}\hline\hline
\text{Label} & $\left(  c_{s}^{2}, Q_{S}, \Omega_{DE}, \omega_{DE}, q \right)
$ & Physical interpretation\\\hline
$P_{1}$ & $\Big(
-\frac{1}{3},3,0, \frac{\left(  4 a_{1} \lambda^{2}+b_{1} \left(  9-6
\lambda^{2}\right)  \right)  ^{2} \tau}{81 b_{1}^{2} (3 b_{1}-2 a_{1})
\lambda}+\mathcal{O}(\tau^{-1}), \frac{\left(  4 a_{1} \lambda^{2}+b_{1}
\left(  9-6 \lambda^{2}\right)  \right)  ^{2} \tau}{54 b_{1}^{2} (3 b_{1}-2
a_{1}) \lambda}+\frac{1}{2}+\mathcal{O}(\tau^{-1})\Big)$ & $a(t)\approx (b_{2}-b_{1}t)^{3b_{1}/2}\left(a_{0}+\mathcal{O}\left(  t^{-1})\right)
\right)  $.\\
& $a_1= -\frac{2 \lambda  \left(\lambda -2 \alpha  \left(\lambda ^2-3\right)\right)\pm\sqrt{6} \sqrt{\lambda  \left(\alpha  \left(6 \alpha  \lambda -2 \lambda
^2+3\right)+\lambda \right)}}{6 \lambda  \left(\alpha  \left(2 \lambda ^2-3\right)-\lambda \right)}, b_1=\frac{2}{9},$ or  & The Galileon mimics radiation for $b_1=\frac{2}{9}.$ \\
& $a_1=\frac{1}{3}, b_1=\frac{4}{9}$, or & The Galileon mimics matter for $b_1=\frac{4}{9}$. \\
& $b_1=\frac{4}{3} a_1, a_1 \left(a_1 \lambda  \left(\alpha  \left(2 \lambda ^2-15\right)-\lambda \right)+1\right)=0$ & Powerlaw-solution for $b_1=\frac{4}{3} a_1\neq 0.$ \\
& & The Galileon mimics dust for \\
& & $  4 a_{1} \lambda^{2}+b_{1} \left(  9-6
\lambda^{2}\right)=0$. \\
& & $c_s^2<0$.\\
$P_{2,3}$ & $\left(  -\frac{1}{3},3,\frac{6}{\lambda^{2}},0,\frac{1}{2}\right)
$ & The Galileon mimics dust.\\
&  & $c_{s}^{2}< 0$.\\
$P_{4,5} $ & $\left(  -\frac{1}{9},-9,1,-1,-1\right)  $ & de Sitter solution.\\
&  & $c_{s}^{2}< 0, Q_{S}<0$.\\
$P_{6,7}(x_{c})$ & \text{see section \ref{discussion-points}} & Accelerated
solution for\\
&  & $x_{c}<0, \lambda<-\frac{3 x_{c}^{2}+2}{\sqrt{6} x_{c}}$, or \\
&  & $x_{c}>0, \lambda>-\frac{3 x_{c}^{2}+2}{\sqrt{6} x_{c}}$\\
$P_{8,9}$ & $\left(  -\frac{4}{9},-9,0,0,\frac{1}{2}\right)  $ & The Galileon
mimics dust.\\
&  & $c_{s}^{2}< 0, Q_{S}<0$.\\
$P_{10,11}$ & $\left(  \frac{2}{9}-\frac{19}{18 \left(  \lambda^{2}-2\right)
},\frac{18}{\lambda^{2}}-9,\frac{3}{2 \lambda^{2}},0,\frac{1}{4} \left(
\frac{9}{\lambda^{2}}+2\right)  \right)  $ & The Galileon mimics dust.\\
&  & $c_{s}^{2}\geq0, Q_{S}>0$ for $0<\lambda^{2}<2$.\\
$P_{12,13}$ & $\left(  \frac{2 (\lambda-1) (\lambda+1) \left(  \lambda
^{2}-6\right)  }{11 \lambda^{4}-18 \lambda^{2}-36},-\frac{3 \left(  11
\lambda^{4}-18 \lambda^{2}-36\right)  }{\left(  \lambda^{2}-6\right)  ^{2}},1,-\frac{3}{2 \lambda^{2}},\frac{1}{2}\right)  $ & The Galileon mimics
dust\\
&  & $c_{s}^{2}\geq0, Q_{S}>0$ for\\
&  & $1\leq\lambda^{2}<\frac{3}{11} \left(  3+\sqrt{53}\right)  \approx
2.80367$.\\
$P_{14}(z_{c})$ & $\left(  -\frac{1}{9},-9,1,-1,-1\right)  $ & de Sitter
solution.\\
& & $\dot\phi \approx 0, a(t)\approx a_0 e^{\frac{t z_c}{\sqrt{1-{z_c}^2}}}, z_c=\pm \sqrt{\frac{V_0}{3+V_0}}.$\\
&  & $c_{s}^{2}< 0, Q_{S}<0$.\\
$P_{15,16}(z_{c})$ & \text{see section \ref{discussion-points}} & de Sitter
solution.\\
& & $\Delta\phi\approx \frac{\sqrt{6} \beta  {z_c} t}{\sqrt{1-{z_c}^2}}, a(t)\approx a_0 e^{\frac{t z_c}{\sqrt{1-{z_c}^2}}}, z_c=\pm \frac{\sqrt{V_0}}{\sqrt{3 \sqrt{6} \alpha\beta +V_0}}.$\\
\hline\hline
\end{tabular}
}\end{table}

We proceed with the determination of critical points at the infinity.

\subsection{Analysis at \textquotedblleft infinity\textquotedblright}

Because the phase space of the system is unbounded we introduce the
Poincar\`{e} compactification and a new time derivative $f^{\prime}$
\begin{equation}
X=\frac{x}{\sqrt{1+x^{2}+y^{2}}}, \quad Y=\frac{y}{\sqrt{1+x^{2}+y^{2}}},
\quad f^{\prime}\rightarrow(1-X^{2}-Y^{2})f^{\prime}.
\end{equation}
The dynamics on the ``cylinder at infinity'' can be obtained by setting
$X=\cos\theta, Y=\sin\theta$; the dynamics in the coordinates $(\theta,z)$ is
governed by the equations%

\begin{subequations}
\begin{align}
&  \theta^{\prime}=h_{1}(\theta,z):=\lambda^{2} z \sin(\theta) \cos^{3}%
(\theta),\label{eq41a}\\
&  z^{\prime}=h_{2}(\theta,z):=\lambda^{2} \left(  z^{2}-1\right)  \cos
^{2}(\theta).
\end{align}

We linearize around a given fixed point on the \textquotedblleft cylinder at
infinity\textquotedblright\ by introducing $X=\cos\theta-\varepsilon
_{1},Y=\sin\theta-\varepsilon_{2}$, with $\varepsilon_{1}\ll1,\varepsilon
_{2}\ll1$. Notice that $1-X^{2}-Y^{2}\simeq2\varepsilon_{1}\cos\theta
+2\varepsilon_{2}\sin\theta$, so to examine the stability of the fixed points
at the cylinder, and from the interior of it, we have to estimate how
$r=\varepsilon_{1}\cos\theta+\varepsilon_{2}\sin\theta$ evolves, not only the
stability in the plane $(\theta,z)$. We obtain for $\varepsilon_{1}%
\ll1,\varepsilon_{2}\ll1$ the expansion rate%

\end{subequations}
\begin{equation}
\label{eq17}r^{\prime}=r\left[  \lambda^{2} z \cos^{2}(\theta) (\cos(2
\theta)-3)\right] .
\end{equation}

\begin{table}[th]
\caption{Summary of the stability conditions of the fixed points at infinity
of the system \eqref{eq.38}.}%
\label{tab:vvvvv}
\centering%
\begin{tabular}
[c]{ccccc}\hline\hline
\text{Label}: \text{Coordinates} $(\theta,z)$ & \text{Coordinates} $(X,Y,z)$ &
$r^{\prime}/r$ & $(\lambda_{1},\lambda_{2})$ & Stability\\\hline
$Q_{1}: \left(  \frac{\pi}{2},z_{c}\right)  $ & $(0,1,z_{c})$ & $0$ & $(0,0)$
& Nonhyperbolic\\
$Q_{2}: (0,-1)$ & $(1,0,-1)$ & $2\lambda^{2}$ & $\left(  -2\lambda
^{2},-\lambda^{2}\right)  $ & Saddle\\
$Q_{3}: (\pi,-1)$ & $(-1,0,-1)$ & $2\lambda^{2}$ & $\left(  -2\lambda
^{2},-\lambda^{2}\right)  $ & Saddle\\
$Q_{4}: (0,1)$ & $(1,0,1)$ & $-2\lambda^{2}$ & $\left(  2\lambda^{2},
\lambda^{2}\right)  $ & Saddle\\
$Q_{5}: (\pi,1)$ & $(-1,0,1)$ & $-2\lambda^{2}$ & $\left(  2\lambda^{2},
\lambda^{2}\right)  $ & Saddle\\\hline\hline
\end{tabular}
\end{table}The stability condition of a fixed point along $r$ is then
$r^{\prime}/r<0$. The full stability of the above fixed points is summarized
in the table \ref{tab:vvvvv}.

\subsection{Numerical analysis}

\label{numeran}

Let us complete our analysis by performing some numerical simulations.
Specifically we choose the constants of the model to satisfy the conditions
\[
g_{0}=-\frac{\lambda\left(  \lambda^{2}p-2\right)  }{4(3p-1)\phi_{0}^{2}%
},~V_{0}=\phi_{0}^{2}\left(  \frac{2}{\lambda^{2}}+p(3p-2)\right)  ,
\]
hence, $\alpha=-\frac{\lambda\left(  \frac{2}{\lambda^{2}}+p(3p-2)\right)
\left(  \lambda^{2}p-2\right)  }{4(3p-1)}.$

\begin{figure}[ht!]
\includegraphics[width=0.35\textwidth]{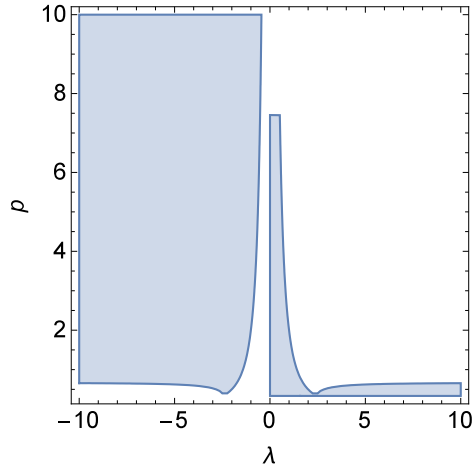}
\caption{``Allowed'' region on the parameter space.}%
\label{Fig:region}%
\end{figure}

Moreover we impose the condition $p>\frac{1}{3}$, which guarantees the
stability of the perturbation of the scaling solution \cite{Dimakis:2017kwx}.
Because we have chosen $\alpha\geq0$, this leads to the \textquotedblleft
allowed\textquotedblright\ region on the parameter space defined by
\begin{enumerate}
\item[i)] $\lambda <-\sqrt{6}, p\geq \frac{1}{3} \sqrt{\frac{\lambda ^2-6}{\lambda ^2}}+\frac{1}{3}$, or
\item[ii)] $\lambda =-\sqrt{6}, p>\frac{1}{3}$, or
\item[iii)] $-\sqrt{6}<\lambda <0, p\geq \frac{2}{\lambda ^2}$, or
\item[iv)] $0<\lambda <\sqrt{6}, \frac{1}{3}<p\leq \frac{2}{\lambda ^2}$, or
\item[v)] $\lambda >\sqrt{6}, \frac{1}{3}<p\leq \frac{1}{3} \sqrt{\frac{\lambda ^2-6}{\lambda ^2}}+\frac{1}{3}$,
\end{enumerate}
as displayed in figure
\ref{Fig:region}.

\begin{figure}[h]
\includegraphics[width=0.8\textwidth]{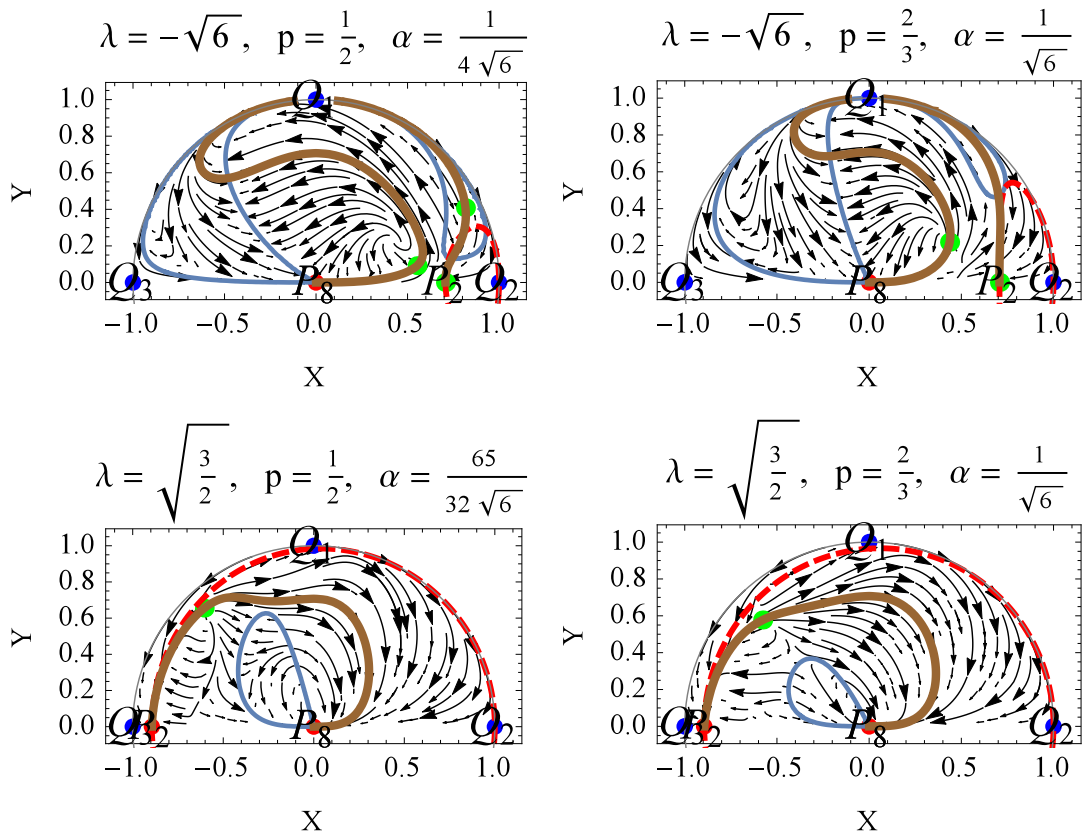} \caption{Poincar\`{e}
projection of the system \eqref{eq.38} on the invariant set $z=-1$. The green
dot corresponds to the points $P_{6}(x_{c})$. In the special case $\lambda
^{2}=6, p=\frac{2}{3}$ we have $1+\alpha\lambda=0$, thus, the polynomial
$\mathcal{P}(x)$ is quadratic and there are only two roots of $\mathcal{P}%
(x)=0$. The blue contour is defined by $f_{4}(x,y,-1)=0$. As shown in the
figures this line is singular and attracts some orbits. The brown solid line
corresponds to intersection of the invariant surface $2\alpha x^{3}\left(
\lambda x-\sqrt{6}z\right)  +y(x^{2}-z^{2})+y^{2}=0$ and the invariant set
$z=-1$. In the top figures, $P_{8}$ attracts some orbits, but others are
attracted by one the of green points associated to $P_{6}(x_{c})$. $P_{8}$ is
not the attractor of the whole phase space since $\alpha\lambda<\frac{1}{4}$.
In the bottom figures $P_{8}$ is the attractor, not just in this invariant set
but in the whole phase space since $\alpha\lambda>\frac{1}{4}$. The red thick
dashed line denotes the local center manifold of $P_{2}$. }%
\label{fig:ProblemCFig1}%
\end{figure}

In figure \ref{fig:ProblemCFig1}, it is presented a Poincar\`{e} projection of
the system \eqref{eq.38} on the invariant set $z=-1$. The green dots
correspond to the points $P_{6}(x_{c})$ (that we solved numerically). In the
special case $\lambda^{2}=6, p=\frac{2}{3}$ we have $1+\alpha\lambda=0$, thus,
the polynomial $\mathcal{P}(x)$ is quadratic and there are only two roots of
$\mathcal{P}(x)=0$. The blue contour is defined by $f_{4}(x,y,-1)=0$. As shown
in the figures this line is singular, and attracts some orbits. The brown
solid line corresponds to intersection of the invariant surface $2\alpha
x^{3}\left(  \lambda x-\sqrt{6}z\right)  +y(x^{2}-z^{2})+y^{2}=0$ and the
invariant set $z=-1$. In the top figures, $P_{8}$ attracts some orbits, but
others are attracted by one the of green points associated to $P_{6}(x_{c})$.
$P_{8}$ is not the attractor of the whole phase space since $\alpha
\lambda<\frac{1}{4}$. In the bottom figures $P_{8}$ is the attractor, not just
in this invariant set but in the whole phase space since $\alpha\lambda
>\frac{1}{4}$.

\begin{figure}[ptb]
\includegraphics[width=0.5\textwidth]{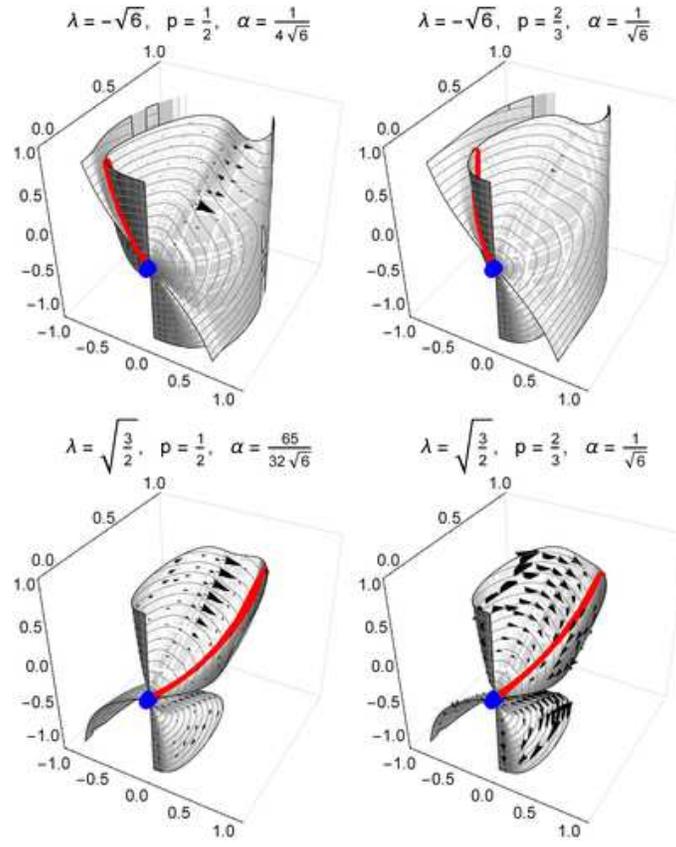}  \caption{Poincar\`{e}
projection of the system \eqref{eq.38} on the invariant surface
\eqref{restr_1} in the coordinates $\left( X,Y, z\right) =\left( \frac
{x}{\sqrt{1+x^{2}+y^{2}}},\frac{y}{\sqrt{1+x^{2}+y^{2}}}, z\right) $. The red
continuous line corresponds to the exact solution \eqref{eq.54}. The origin
(represented by a blue dot) attracts this line. The vector field \eqref{eq.38}
is projected onto the surface.}%
\label{fig:Fig2}%
\end{figure}

\subsection{Discussion}

\label{discussion-points}

In this section we discuss the stability conditions, cosmological properties
and physical meaning of the (lines of) fixed points in both finite and
infinite regions.

\begin{itemize}
\item $P_{1}: \left(  0, 0, 0 \right)  $ always exists. To analyze the
stability we resort to numerical examination.

In the Appendix \ref{appendixA} we proved, using normal forms calculations,
that the fixed point $P_{1}$ corresponds to the cosmological solution
\eqref{EqB15}.\ Moreover, in order to improve the range and accuracy we
calculate the diagonal first order Pade approximants
\[
\lbrack1/1]_{\dot{\phi}}(t),\quad\lbrack1/1]_{H}(t),\quad\lbrack1/1]_{\phi
}(t),
\]
around $t=\infty$. This yields the following approximate expressions
\begin{subequations}
\begin{align}
& \dot\phi(t)\approx\frac{1}{4} \lambda(2 a_{1}-3 b_{1}) \left( \frac{8
a_{1}-12 b_{1}}{-2 a_{1} t+2 a_{2}+3 b_{1} t-3 b_{2}}+\frac{3 b_{1} \ln
t}{t^{2}}\right) ,\\
&  H(t)\approx\frac{3}{8} b_{1}^{2} \left( -\frac{3 \ln t}{t^{2}}-\frac
{4}{b_{2}-b_{1} t}\right) ,\\
&  \phi(t)\approx\frac{\ln\left( \frac{4 V_{0}}{4 \alpha\lambda^{3} (2 a_{1}-3
b_{1})^{2}+2 \lambda^{2} (2 a_{1}-3 b_{1})^{2}+3b_{1} (9 b_{1}-4)}\right)
}{\lambda}+\frac{2 \ln t}{\lambda},
\end{align}
while the scale factor is calculated to be
\end{subequations}
\begin{equation}
\label{scalar-factor-P1}a(t)\approx a_{0}e^{\frac{9b_{1}^{2}}{8t}}\left(
\frac{1}{t}\right)  ^{-\frac{9b_{1}^{2}}{8t}}(b_{2}-b_{1}t)^{3b_{1}/2}%
=(b_{2}-b_{1}t)^{3b_{1}/2}\left(  a_{0}+\mathcal{O}\left(  t^{-1})\right)
\right) .
\end{equation}
Due to the new conservation law \eqref{con01}, the allowed values of the
constants $a_{1}, a_{2}, b_{1}, b_{2}$ are:

\begin{itemize}
\item[i)]
\begin{align*}
&  a_{1}= -\frac{2 \lambda\left( \lambda-2 \alpha\left( \lambda^{2}-3\right)
\right) \pm\sqrt{6} \sqrt{\lambda\left( \alpha\left( 6 \alpha\lambda-2
\lambda^{2}+3\right) +\lambda\right) }}{6 \lambda\left( \alpha\left( 2
\lambda^{2}-3\right) -\lambda\right) }, \quad b_{1}=\frac{2}{9},\\
&  {a_{1}} (6-162 {b_{2}})+36 {a_{2}}-1=0, \quad4 {a_{1}}-\frac{2 }{3}%
+\frac{I_{1}}{{a_{0}}^{3}}=0, \quad\text{or}%
\end{align*}

\item[ii)]
\[
a_{1}=\frac{1}{3}, \quad b_{1}=\frac{4}{9}, \quad2 \alpha\lambda^{3}-15
\alpha\lambda-\lambda^{2}=0, \quad\frac{I_{1}}{{a_{0}}^{3}}+4 a_{2}-3 b_{2}=0,
\quad\text{or}%
\]

\item[iii)]
\[
a_{1} \left( a_{1} \lambda\left( \alpha\left( 2 \lambda^{2}-15\right)
-\lambda\right) +1\right) =0, \quad b_{1}=\frac{4}{3} a_{1}, \quad b_{2}%
=\frac{4}{3} a_{2}, \quad I_{1}=0.
\]

\end{itemize}

It is interesting to note that the power-law solution (\ref{solution1})
satisfies the condition
\begin{equation}
(x\left(  \tau\right)  ,y\left(  \tau\right)  ,z\left(  \tau\right)  )=\left(
\frac{\sqrt{\frac{2}{3}}}{\lambda\sqrt{p^{2}+t^{2}}},\quad\frac{V_{0}}%
{3\phi_{0}^{2}\left(  p^{2}+t^{2}\right)  },\quad\frac{p}{\sqrt{p^{2}+t^{2}}%
}\right)  . \label{eq.54}%
\end{equation}
Additionally, after the substitution of the functional forms of $(x\left(
\tau\right)  ,y\left(  \tau\right)  ,z\left(  \tau\right)  )$ in
\eqref{eq.54}, and the substitution of $g_{0}=-\frac{\lambda\left(
\lambda^{2}p-2\right)  }{4(3p-1)\phi_{0}^{2}},~V_{0}=\phi_{0}^{2}\left(
\frac{2}{\lambda^{2}}+p(3p-2)\right)  $ it follows that the restriction
(\ref{eq.54}) is satisfied for all the values of $t$.

There exists a relation between $\tau$ and $t$ given by
\begin{equation}
\tau=\sqrt{p^{2}+t^{2}}-p\ln\left(  \frac{p\left(  \sqrt{p^{2}+t^{2}%
}+p\right)  }{t}\right)  ,
\end{equation}
such that $t\rightarrow\infty$ implies $\tau\rightarrow\infty$. Thus, as
$\tau\rightarrow\infty$, this power-law solution approaches the $P_{1}$ as
$t\rightarrow\infty$. For large $\tau$ we can invert to have
\begin{equation}
t=\frac{1}{4}\left(  \sqrt{8p^{2}+4(p\ln(p)+\tau)^{2}}+2p\ln(p)+2\tau\right)
=\tau+p\ln(p)+\frac{p^{2}}{2\tau}-\frac{p^{3}\ln(p)}{2\tau^{2}}+\mathcal{O}%
\left(  \tau^{-3}\right)  .
\end{equation}
Thus, we can take as approximation for large $\tau$:
\[
x=\frac{\sqrt{\frac{2}{3}}}{\lambda\tau}-\frac{\sqrt{\frac{2}{3}}p\ln
(p)}{\lambda\tau^{2}}+\mathcal{O}\left(  \tau^{-3}\right)  ,\quad
y=\frac{V_{0}}{3\phi_{0}^{2}\tau^{2}}+\mathcal{O}\left(  \tau^{-3}\right)
,\quad z=\frac{p}{\tau}-\frac{p^{2}\ln(p)}{\tau^{2}}+\mathcal{O}\left(
\tau^{-3}\right)  .
\]
Furthermore
\[
\lambda\phi=\left(  2\ln(\phi_{0})+2\ln\tau\right)  +\frac{2p\ln(p)}{\tau
}-\frac{p^{2}\left(  \ln^{2}(p)-1\right)  }{\tau^{2}}+\mathcal{O}\left(
\tau^{-3}\right)  ,\quad\dot{\phi}=\frac{2}{\lambda\tau}-\frac{2(p\ln
(p))}{\lambda\tau^{2}}+\mathcal{O}\left(  \tau^{-3}\right)  .
\]
These features are represented in Figure \ref{fig:Fig2}. There it is shown the
Poincar\`{e} projection of the system \eqref{eq.38} on the invariant surface
\eqref{restr_1} in the variables $\left( X,Y,z\right) =\left( \frac{x}%
{\sqrt{1+x^{2}+y^{2}}},\frac{y}{\sqrt{1+x^{2}+y^{2}}},z\right) $. The red
continuous line corresponds to the exact solution \eqref{eq.54}. The origin
(represented by a blue dot) attracts this line. The vector field \eqref{eq.38}
is projected onto the surface.

The values of $\left(  c_{s}^{2},Q_{S},\Omega_{DE},\omega_{DE},q\right)  $ for
$P_{1}$ are $\Big(-\frac{1}{3},3,0,\frac{\left(  4a_{1}\lambda^{2}%
+b_{1}\left(  9-6\lambda^{2}\right)  \right)  ^{2}\tau}{81b_{1}^{2}%
(3b_{1}-2a_{1})\lambda}+\mathcal{O}(\tau^{-1}),\frac{\left(  4a_{1}\lambda
^{2}+b_{1}\left(  9-6\lambda^{2}\right)  \right)  ^{2}\tau}{54b_{1}^{2}%
(3b_{1}-2a_{1})\lambda}+\frac{1}{2}+\mathcal{O}(\tau^{-1})\Big)$. The scale
factor satisfy $a(t)\approx(b_{2}-b_{1}t)^{3b_{1}/2}\left(  a_{0}%
+\mathcal{O}\left(  t^{-1}\right)  \right) $. The Galileon mimics radiation
for $b_{1}=\frac{2}{9}.$ It mimics matter for $b_{1}=\frac{4}{9}$. It is a
Powerlaw-solution for $b_{1}=\frac{4}{3} a_{1}\neq0.$ Finally, the Galileon
mimics dust for $4 a_{1} \lambda^{2}+b_{1} \left(  9-6 \lambda^{2}\right) =0$.
Furthermore $c_{s}^{2}<0$. This point has not been obtained previously in
\cite{genlyGL} and in \cite{genlyGL2}, since in these works the authors used
$H$-normalization, which fails obviously when $H=0$.
\item $P_{2}:\left(  -\frac{\sqrt{6}}{\lambda},0,-1\right)  $ always exists.
The eigenvalues are $0,-3,-6$, so, the points are nonhyperbolic. Using the
Center Manifold Theory we have proven that $P_{2}$ is stable for
$\lambda<-\sqrt{6}$ or $0<\lambda<\sqrt{6}$, and saddle for $-\sqrt{6}%
<\lambda<0$ or $\lambda>\sqrt{6}$. (see appendix \ref{AppP2}). This point
represents kinetic dominated solutions with $H\rightarrow-\infty$. The values
of $\left(  c_{s}^{2},Q_{S},\Omega_{DE},\omega_{DE},q\right)  $ for $P_{2}$
are $\left(  -\frac{1}{3},3,\frac{6}{\lambda^{2}},0,\frac{1}{2}\right)  $. The
Galileon mimics dust. Furthermore $c_{s}^{2}<0$. So, the Laplacian
instabilities associated with the scalar field propagation speed cannot be
avoided for this solution \cite{DeFelice:2011bh}.

\item $P_{3}: \left(  \frac{\sqrt{6}}{\lambda}, 0, 1 \right)  $ always exists.
The eigenvalues are $0, 3, 6$, so, the points are nonhyperbolic. This point
has the opposite dynamical behavior of $P_{2}$. Using the Center Manifold
Theory (in a similar way as in the appendix \ref{AppP2} we can prove that
$P_{3}$ is unstable for $\lambda<-\sqrt{6}$ or $0<\lambda<\sqrt{6}$, and
saddle for $-\sqrt{6}<\lambda<0$ or $\lambda>\sqrt{6}$). This point represents
kinetic dominated solutions with $H\rightarrow+\infty$. The values of $\left(
c_{s}^{2}, Q_{S}, \Omega_{DE}, \omega_{DE}, q \right)  $ for $P_{3}$ are
$\left(  -\frac{1}{3},3,\frac{6}{\lambda^{2}},0,\frac{1}{2}\right)  $. The
Galileon mimics dust. Furthermore $c_{s}^{2}< 0$. So, the Laplacian
instabilities associated with the scalar field propagation speed cannot be
avoided for this solution \cite{DeFelice:2011bh}.

\item The fixed point $P_{4}:\left(  0,1,1\right)  $ exists if $\lambda=0$. As
before, it corresponds to the special case $V=V_{0}$, and the coupling
function becomes constant too, $g=g_{0}$; it is a de Sitter solution, but now
$H\rightarrow+\infty$. The eigenvalues are $-3,-3,0$, so it is nonhyperbolic.
Using the Center Manifold Theory we obtain that it is asymptotically stable
(for details see Appendix \ref{AppP4}). Furthermore, perturbations from the
equilibrium grow or decay algebraically in time, not exponentially as in the
usual linear stability analysis. The values of $\left(  c_{s}^{2},Q_{S}%
,\Omega_{DE},\omega_{DE},q\right)  $ for $P_{4}$ are $\left(  -\frac{1}%
{9},-9,1,-1,-1\right)  $. As it can be seen, $c_{s}^{2}<0,Q_{S}<0$. Thus, this
solution suffers from Laplacian instabilities and the presence of ghosts.

\item The fixed point $P_{5}:\left(  0,1,-1\right)  $ exists if $\lambda=0$.
The eigenvalues are $3, 3, 0$, so it is nonhyperbolic. To analyze their
stability we resort to numerical examination or use the Center Manifold
Theory. This fixed point corresponds to a de Sitter solution driven by a
cosmological constant, since $V=V_{0}$, and the coupling function becomes
constant too, $g=g_{0}$ (although, $H\rightarrow-\infty$, so, it is not
cosmological viable). This point has the opposite dynamical behavior of
$P_{4}$, thus it is asymptotically unstable. The values of $\left(  c_{s}^{2},
Q_{S}, \Omega_{DE}, \omega_{DE}, q \right)  $ for $P_{5}$ are $\left(
-\frac{1}{9},-9,1,-1,-1\right)  $. The associated cosmological solution
corresponds to a de Sitter solution. However, $c_{s}^{2}< 0, Q_{S}<0$. Thus,
this solution suffers from Laplacian instabilities and the presence of ghosts.

\item For each choice $\epsilon=\pm1$, there are 1, 2 or 3 isolated fixed
points of the form $P_{6,7}(x_{c}): \left(  \epsilon x_{c}, \sqrt{\frac{2}{3}}
\lambda x_{c}+x_{c}^{2}+1, -\epsilon\right)  $, where $x_{c}$ are the nonzero
real roots of the polynomial $\mathcal{P}(x)=6(1+\alpha\lambda) {x}^{3}%
+3\sqrt{6}(2\alpha+\lambda){x}^{2} +2(3+\lambda^{2})x +\sqrt{6}\lambda$,
satisfying $\sqrt{\frac{2}{3}}\lambda x_{c}+x_{c}^{2}+1\geq0$. To analyze
their stability we resort to numerical examination.

The values of $\left(  c_{s}^{2}, Q_{S}, \Omega_{DE}, \omega_{DE}, q \right) $
for $P_{6,7}(x_{c})$ are ${c_{s}^{2}}^{*}=-P_{1}(x_{c})/(Q_{1}(x_{c})
Q_{2}(x_{c}))$, $P_{1}(x_{c})=144 \alpha^{4} x_{c}^{10}+24 \alpha^{3}
x_{c}^{7} y_{c} \left(  \lambda x_{c}+2 \sqrt{6}\right)  +6 \alpha^{2}
x_{c}^{4} y_{c}^{2} \left(  2 \left(  \lambda^{2}-7\right)  x_{c}^{2}+6
\sqrt{6} \lambda x_{c}+18 y_{c}+11\right)  +2 \alpha x_{c} y_{c}^{3} \left(  3
\lambda x_{c}^{3}+\lambda x_{c} \left(  21 y_{c}-8\right)  +12 \sqrt{6}
x_{c}^{2}+\sqrt{6} \left(  3 y_{c}-4\right)  \right)  +y_{c}^{4} \left(  -3
x_{c}^{2}+3 y_{c}-4\right)  $, $Q_{1}(x_{c})= 3 \left(  6 \alpha^{2} x_{c}%
^{4}+2 \alpha x_{c} y_{c} \left(  2 \lambda x_{c}+\sqrt{6}\right)  +y_{c}%
^{2}\right)  $, and $Q_{2}(x_{c})=24 \alpha^{2} x_{c}^{6}-4 \alpha x_{c}^{3}
y_{c} \left(  \sqrt{6}-4 \lambda x_{c}\right)  +\left(  4 x_{c}^{2}-3\right)
y_{c}^{2}$, where $y_{c}=\sqrt{\frac{2}{3}}\lambda x_{c}+x_{c} ^{2}+1\geq0$.
$Q_{S}^{*}= \frac{9 \left(  12 \left(  6 \alpha^{2}+4 \alpha\lambda+1\right)
x_{c}^{6}+4 \sqrt{6} \left(  \alpha\left(  4 \lambda^{2}-3\right)  +2
\lambda\right)  x_{c}^{5}+(8 \lambda(3 \alpha+\lambda)+15) x_{c}^{4}+2
\sqrt{6} (\lambda-6 \alpha) x_{c}^{3}-6 \left(  \lambda^{2}+1\right)
x_{c}^{2}-6 \sqrt{6} \lambda x_{c}-9\right)  }{\left(  -6 \sqrt{6} \alpha
x_{c}^{3}+\sqrt{6} \lambda x_{c}+3 x_{c}^{2}+3\right)  {}^{2}}$, $\Omega
_{DE}^{*}=\frac{x_{c} \left(  x_{c} \left(  6 (\alpha\lambda+1) x_{c}^{2}+3
\sqrt{6} (2 \alpha+\lambda) x_{c}+2 \lambda^{2}+9\right)  +2 \sqrt{6}
\lambda\right)  +3}{\sqrt{6} \lambda x_{c}+3 x_{c}^{2}+3}$, $\omega_{DE}%
^{*}=-\frac{1}{3} x_{c} \left(  3 x_{c}+\sqrt{6} \lambda\right)  -1$, and
$q^{*}=\frac{1}{2} \left(  -x_{c} \left(  3 x_{c}+\sqrt{6} \lambda\right)
-2\right) $.

\begin{figure}[ht!]
\centering
\includegraphics[width=0.5\textwidth]{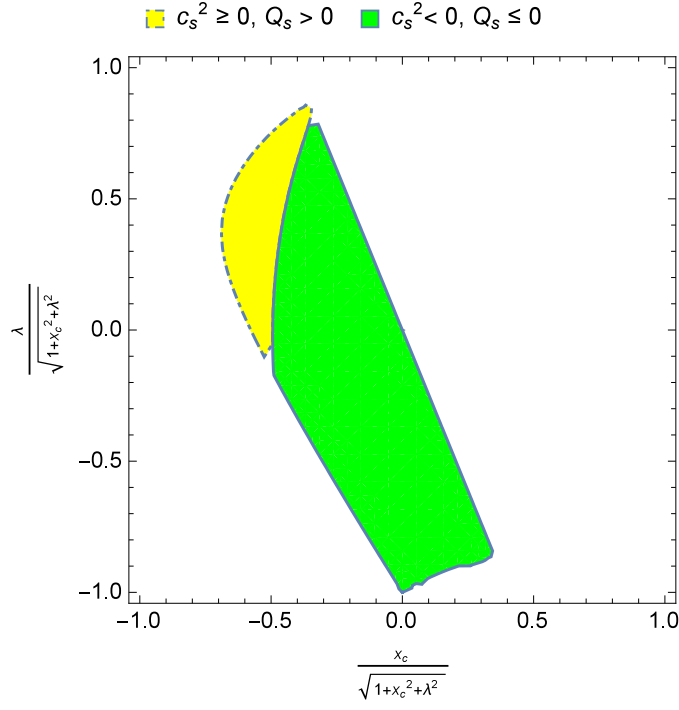}  \caption{Parameter space
that leads to the conditions $c_{s}^{2}\geq0, Q_{S}>0$ and $c_{s}^{2}< 0,
Q_{S}\leq0$, for $P_{6, 7}(x_{c})$.}%
\label{fig:RegionP6}%
\end{figure}

The fixed points
$P_{6,7}(x_{c})$ represents accelerated solution for $x_{c}<0, \lambda
<-\frac{3 x_{c}^{2}+2}{\sqrt{6} x_{c}}$, or $x_{c}>0, \lambda>-\frac{3
x_{c}^{2}+2}{\sqrt{6} x_{c}}$.
Since the above analytical expressions for $c_{s}^2$ and $Q_{S}$
are quite complicated, we have resorted to numerical investigation. To represent the regions of physical interest, we proceed in the following way. Recall that the polynomial $\mathcal{P}(x)=6(1+\alpha\lambda) {x}^{3}%
+3\sqrt{6}(2\alpha+\lambda){x}^{2} +2(3+\lambda^{2})x +\sqrt{6}\lambda$ has a discrete number of roots $x_c$ (1, 2, or 3 depending on the parameters $\alpha$ and $\lambda$). Since this polynomial is linear in $\alpha$, we have for each value of $x_c$ ($x_c\neq 0, x_c\neq -\frac{\sqrt{6}}{\lambda}$) the relation $\alpha:=\alpha(x_c, \lambda)=-\frac{6 \left(x_{c}^3+x_{c}\right)+\sqrt{6} \lambda  \left(3 x_{c}^2+1\right)+2 \lambda ^2 x_{c}}{6 x_{c}^2 \left(\lambda  x_{c}+\sqrt{6}\right)}, \alpha>0$. So, we can represent the regions of physical interests on the parameter space $(x_c, \lambda)$, rather than in the plane $(\alpha, \lambda)$ (to avoid solving a generically third order polynomial in $x_c$ using Cardano's formulas, with the subsequent numerical errors issue). The above procedure leads to the conditions
$c_{s}^{2}\geq0, Q_{S}>0$ and $c_{s}^{2}< 0, Q_{S}\leq0$, for $P_{6,7}(x_{c})$ as displayed in Figure \ref{fig:RegionP6}. In order to cover all the possible values of the parameters we have made the representation in the (compact) variables $\left(\frac{x_c}{\sqrt{1+x_c^2+\lambda^2}},\frac{\lambda}{\sqrt{1+x_c^2+\lambda^2}}\right)$.

\item The fixed point $P_{8}:\left(  0,0,-1\right)  $ always exists. The
eigenvalues are $-3,-3, -\left(  6 \alpha\lambda-\frac{3}{2}\right)  $. Thus,
it is a sink for $\alpha\lambda>\frac{1}{4}$ or saddle otherwise. This
solution is dominated by the Hubble scalar with $H\rightarrow-\infty$. The
values of $\left(  c_{s}^{2}, Q_{S}, \Omega_{DE}, \omega_{DE}, q \right)  $
for $P_{8}$ are $\left(  -\frac{4}{9},-9,0,0,\frac{1}{2}\right)  $. The
Galileon mimics dust. However, the fluid satisfies the conditions $c_{s}%
^{2}<0, Q_{S}<0$. Thus, this solution suffers from Laplacian instabilities and
the presence of ghosts.

\item The fixed point $P_{9}:\left(  0,0,1\right)  $ always exists. The
eigenvalues are $3,3, \left(  6 \alpha\lambda-\frac{3}{2}\right)  $. Thus, it
is a source for $\alpha\lambda>\frac{1}{4}$ or a saddle otherwise. This
solution is dominated by the Hubble scalar with $H\rightarrow+\infty$. The
values of $\left(  c_{s}^{2}, Q_{S}, \Omega_{DE}, \omega_{DE}, q \right)  $
for $P_{9}$ are $\left(  -\frac{4}{9},-9,0,0,\frac{1}{2}\right)  $. The
Galileon mimics dust. However, the fluid satisfies the conditions $c_{s}%
^{2}<0, Q_{S}<0$. Thus, this solution suffers from Laplacian instabilities and
the presence of ghosts.

\item The fixed point $P_{10}: \left(  -\frac{\sqrt{\frac{3}{2}}}{\lambda}, 0
, -1\right)  $ exists for $\alpha=0, \lambda\neq0$. The eigenvalues are
$\left(  -\frac{9}{2 \lambda^{2}},\frac{3}{4} \left(  2-\frac{9}{\lambda^{2}%
}\right)  ,-\frac{9}{2 \lambda^{2}}-3\right)  $. Thus, it is a sink for
$0<\lambda^{2}<\frac{3}{2}$ and a saddle otherwise. The values of $\left(
c_{s}^{2}, Q_{S}, \Omega_{DE}, \omega_{DE}, q \right)  $ for $P_{10}$ are
$\left(  \frac{2}{9}-\frac{19}{18 \left(  \lambda^{2}-2\right)  },\frac
{18}{\lambda^{2}}-9,\frac{3}{2 \lambda^{2}},0,\frac{1}{4} \left(  \frac
{9}{\lambda^{2}}+2\right)  \right)  $. The Galileon mimics dust. Furthermore,
$c_{s}^{2}\geq0, Q_{S}>0$ for $0<\lambda^{2}<2$. In this region of the
parameter space, the cosmological solution is free of Laplacian instabilities
and ghosts.

\item The fixed point $P_{11}: \left(  \frac{\sqrt{\frac{3}{2}}}{\lambda}, 0,
1\right)  $ exists for $\alpha=0, \lambda\neq0$. The eigenvalues are $\left(
\frac{9}{2 \lambda^{2}},-\frac{3}{4} \left(  2-\frac{9}{\lambda^{2}}\right)
,\frac{9}{2 \lambda^{2}}+3\right)  $. It is a source for $0<\lambda^{2}%
<\frac{3}{2}$ or a saddle otherwise. The values of $\left(  c_{s}^{2}, Q_{S},
\Omega_{DE}, \omega_{DE}, q \right)  $ for $P_{10}$ are $\left(  \frac{2}%
{9}-\frac{19}{18 \left(  \lambda^{2}-2\right)  },\frac{18}{\lambda^{2}%
}-9,\frac{3}{2 \lambda^{2}},0,\frac{1}{4} \left(  \frac{9}{\lambda^{2}%
}+2\right)  \right)  $. The Galileon mimics dust. Furthermore, $c_{s}^{2}%
\geq0, Q_{S}>0$ for $0<\lambda^{2}<2$. In this region of the parameter space,
the cosmological solution is free of Laplacian instabilities and ghosts.

\item The fixed point $P_{12}: \left(  -\frac{\sqrt{\frac{3}{2}}}{\lambda},
\frac{3}{2\lambda^{2}}, -1\right)  $ exists for $\lambda^{2}+3\alpha
\lambda-3=0$. The eigenvalues are $0,3,-\frac{3}{2}$. It is a saddle. The
values of $\left(  c_{s}^{2}, Q_{S}, \Omega_{DE}, \omega_{DE}, q \right)  $
for $P_{12}$ are $\left(  \frac{2 (\lambda-1) (\lambda+1) \left(  \lambda
^{2}-6\right)  }{11 \lambda^{4}-18 \lambda^{2}-36},-\frac{3 \left(  11
\lambda^{4}-18 \lambda^{2}-36\right)  }{\left(  \lambda^{2}-6\right)  ^{2}%
},1,-\frac{3}{2 \lambda^{2}},\frac{1}{2}\right)  $. The Galileon mimics dust.
Furthermore, $c_{s}^{2}\geq0, Q_{S}>0$ for $1\leq\lambda^{2}<\frac{3}{11}
\left(  3+\sqrt{53}\right)  \approx2.80367$. In this region of the parameter
space, the cosmological solution is free of Laplacian instabilities and ghosts.

\item The fixed point $P_{13}: \left(  \frac{\sqrt{\frac{3}{2}}}{\lambda},
\frac{3}{2\lambda^{2}}, 1\right)  $ exists for $\lambda^{2}+3\alpha
\lambda-3=0$. The eigenvalues are $0,3,-\frac{3}{2}$. It is a saddle. The
values of $\left(  c_{s}^{2}, Q_{S}, \Omega_{DE}, \omega_{DE}, q \right)  $
for $P_{12}$ are $\left(  \frac{2 (\lambda-1) (\lambda+1) \left(  \lambda
^{2}-6\right)  }{11 \lambda^{4}-18 \lambda^{2}-36},-\frac{3 \left(  11
\lambda^{4}-18 \lambda^{2}-36\right)  }{\left(  \lambda^{2}-6\right)  ^{2}%
},1,-\frac{3}{2 \lambda^{2}},\frac{1}{2}\right)  $. The Galileon mimics dust.
Furthermore, $c_{s}^{2}\geq0, Q_{S}>0$ for $1\leq\lambda^{2}<\frac{3}{11}
\left(  3+\sqrt{53}\right)  \approx2.80367$. In this region of the parameter
space, the cosmological solution is free of Laplacian instabilities and ghosts.

\item The line of fixed points $P_{14}(z_{c}):\left(  0,z_{c}^{2}%
,z_{c}\right)  $ exists for $\lambda=0$. The eigenvalues are $\left(
0,-3z_{c},-3z_{c}\right) $. Thus, it is nonhyperbolic. Using the Center
Manifold Theory we obtain that it is asymptotically stable for $0<{z_{c}}%
\leq1$ and asymptotically unstable for $-1\leq{z_{c}}<0$ (for details see
Appendix \ref{AppP14}). The associated cosmological solution corresponds to a
de Sitter solution that satisfies $\phi, H$ are approximately constant, such
that
\[
\dot\phi\approx0, \quad H(t) \approx\frac{z_{c}}{\sqrt{1-{z_{c}}^{2}}}, \quad
a(t)\approx a_{0} e^{\frac{t z_{c}}{\sqrt{1-{z_{c}}^{2}}}}, z_{c}=\pm
\sqrt{\frac{V_{0}}{3+V_{0}}},
\]
where the potential is constant $V(\phi)=V_{0}$ since $\lambda=0$. The values
of $\left(  c_{s}^{2},Q_{S},\Omega_{DE},\omega_{DE},q\right)  $ for
$P_{14}(z_{c})$ are $\left(  -\frac{1}{9},-9,1,-1,-1\right) $. However,
$c_{s}^{2}<0,Q_{S}<0$. Thus, this solution suffers from Laplacian
instabilities and the presence of ghosts.

\item The line of fixed points $P_{15}(z_{c}):\left(  \beta z_{c},\sqrt
{6}\alpha\beta{z_{c}}^{2},z_{c}\right)  $ exists for $\lambda=0, \beta
=\frac{1}{\sqrt{2}}\left(  \sqrt{3}\alpha-\sqrt{3\alpha^{2}-2}\right) $. The
eigenvalues are $\left(  0,-3z_{c},-3z_{c}\right)  $. Thus, it is
nonhyperbolic. The stability has to be examined numerically or using Center
Manifold calculations. The values of $\left(  c_{s}^{2},Q_{S},\Omega
_{DE},\omega_{DE},q\right)  $ for $P_{15}(z_{c})$ are $c_{s}^{2}=\frac{\left(
\alpha\left( 19 \sqrt{9 \alpha^{2}-6}-3 \alpha\left( 51-8 \alpha\left( 3
\alpha\left( 2 \alpha\left( \sqrt{9 \alpha^{2}-6}-3 \alpha\right) +7\right) -5
\sqrt{9 \alpha^{2}-6}\right) \right) \right) +6\right)  {z_{c}}}{\left(
\alpha\left( \sqrt{9 \alpha^{2}-6}-3 \alpha\right) +2\right)  \left( 3 \left(
24 \left( \alpha\left( \sqrt{9 \alpha^{2}-6}-3 \alpha\right) +1\right)
\alpha^{2}+7\right)  {z_{c}}-8 \left( \sqrt{9 \alpha^{2}-6}-3 \alpha\right)
\sqrt{1-{z_{c}}^{2}}\right) }$, $Q_{S}=\frac{8 \left( \sqrt{9 \alpha^{2}-6}-3
\alpha\right)  \sqrt{1-{z_{c}}^{2}}+3 \left( 24 \alpha^{2} \left( 3 \alpha
^{2}-\sqrt{9 \alpha^{2}-6} \alpha-1\right) -7\right)  {z_{c}}}{\left( 2
\alpha\left( \sqrt{9 \alpha^{2}-6}-3 \alpha\right) +1\right) ^{2} {z_{c}}}$,
$\Omega_{DE}=1$, $\omega_{DE}=\frac{\left( \sqrt{9 \alpha^{2}-6}-3
\alpha\right)  \left( 6 \alpha{z_{c}}-\sqrt{1-{z_{c}}^{2}}\right) }{6 {z_{c}}%
}$, and $q=-1$. This solution is free of Laplacian instabilities and
ghosts-free in the region displayed in figure \ref{fig:RegionP15} (we used the
compact variable $\alpha/(1+\alpha)$ to cover all real values of $\alpha$).

\begin{figure}[h]
\centering
\includegraphics[width=0.4\textwidth]{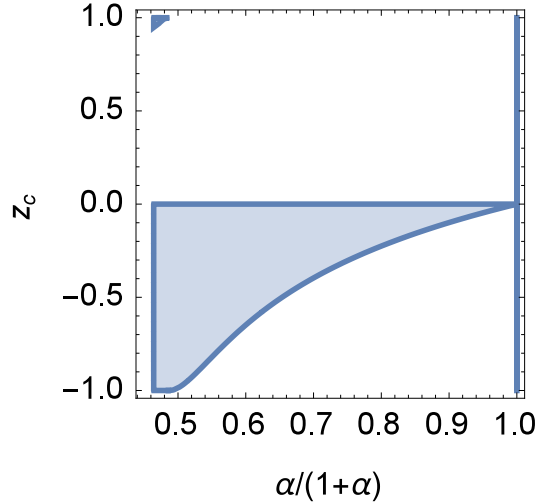}  \caption{Parameter space
where $P_{15}(z_{c})$ is free from Lapacian instabilities and it is ghost
-free.}%
\label{fig:RegionP15}%
\end{figure}

\item The line of fixed points $P_{16}(z_{c}):\left(  \beta z_{c},\sqrt
{6}\alpha\beta{z_{c}}^{2},z_{c}\right)  $ exists for $\lambda=0, \beta
=\frac{1}{\sqrt{2}}\left(  \sqrt{3}\alpha+\sqrt{3\alpha^{2}-2}\right) $. The
eigenvalues are $\left(  0,-3z_{c},-3z_{c}\right) $. Thus, it is
nonhyperbolic. The stability has to be examined numerically or using Center
Manifold calculations. The values of $\left(  c_{s}^{2},Q_{S},\Omega
_{DE},\omega_{DE},q\right) $ for $P_{16}(z_{c})$ are $c_{s}^{2}=\frac{\left(
6-\alpha\left( 3 \alpha\left( 8 \alpha\left( 3 \alpha\left( 2 \alpha\left(
\sqrt{9 \alpha^{2}-6}+3 \alpha\right) -7\right) -5 \sqrt{9 \alpha^{2}%
-6}\right) +51\right) +19 \sqrt{9 \alpha^{2}-6}\right) \right)  {z_{c}}%
}{\left( \alpha\left( \sqrt{9 \alpha^{2}-6}+3 \alpha\right) -2\right)  \left(
3 \left( 24 \alpha^{2} \left( \alpha\left( \sqrt{9 \alpha^{2}-6}+3
\alpha\right) -1\right) -7\right)  {z_{c}}-8 \left( \sqrt{9 \alpha^{2}-6}+3
\alpha\right)  \sqrt{1-{z_{c}}^{2}}\right) }$, $Q_{S}=\frac{3 \left( 24
\alpha^{2} \left( \alpha\left( \sqrt{9 \alpha^{2}-6}+3 \alpha\right) -1\right)
-7\right)  {z_{c}}-8 \left( \sqrt{9 \alpha^{2}-6}+3 \alpha\right)
\sqrt{1-{z_{c}}^{2}}}{\left( 1-2 \alpha\left( \sqrt{9 \alpha^{2}-6}+3
\alpha\right) \right) ^{2} {z_{c}}}$, $\Omega_{DE}=1$, $\omega_{DE}%
=\frac{\left( \sqrt{9 \alpha^{2}-6}+3 \alpha\right)  \left( \sqrt{1-{z_{c}%
}^{2}}-6 \alpha{z_{c}}\right) }{6 {z_{c}}}$, and $q=-1$. The line of fixed
points $P_{16}(z_{c})$ always satisfy $c_{s}^{2} Q_{S}<0$. Thus, these
solutions suffers either from Laplacian instabilities or from the presence of ghosts.

For the lines of fixed points $P_{15}(z_{c})$ and $P_{16}(z_{c})$ the
cosmological solutions satisfy
\[
\dot\phi(t)\approx\frac{\sqrt{6} \beta{z_{c}}}{\sqrt{1-{z_{c}}^{2}}}, \quad
H(t)\approx\frac{{z_{c}}}{\sqrt{1-{z_{c}}^{2}}}, \quad\phi(t)-\phi_{0}%
\approx\frac{\sqrt{6} \beta{z_{c}} t}{\sqrt{1-{z_{c}}^{2}}}, \quad a(t)\approx
a_{0} e^{\frac{t z_{c}}{\sqrt{1-{z_{c}}^{2}}}}, \quad z_{c}=\pm\frac
{\sqrt{V_{0}}}{\sqrt{3 \sqrt{6} \alpha\beta+V_{0}}},
\]
where $\beta=\frac{\sqrt{3}\alpha\mp\sqrt{3\alpha^{2}-2}}{\sqrt{2}}$
respectively, and the potential is constant $V(\phi)=V_{0}$ since $\lambda=0$.
\end{itemize}

Finally, the fixed points/ fixed lines at infinity are

\begin{itemize}
\item $\left(  \frac{\pi}{2}, z_{c}\right)  $ with eigenvalues $(0,0)$. So, it
is non-hyperbolic. According to the analysis in \ref{analysisQ1} it is
generically a saddle.

\item $(0,-1)$ with eigenvalues $\left(  -2\lambda^{2},-\lambda^{2}\right)  $.
This point attracts nearby orbits lying on the cylinder for all $\lambda\neq
0$. However, if we take into account the stability along $r$, the point is
generically a saddle.

\item $(\pi,-1)$ with eigenvalues $\left(  -2\lambda^{2},-\lambda^{2}\right)
$. This point attracts nearby orbits lying on the cylinder for all
$\lambda\neq0$. However, if we take into account the stability along $r$, the
point is generically a saddle.

\item $(0,1)$ with eigenvalues $\left(  2\lambda^{2},\lambda^{2}\right)  $.
This point repels nearby orbits lying on the cylinder for all $\lambda\neq0$.
However, if we take into account the stability along $r$, the point is
generically a saddle.

\item $(\pi,1)$ with eigenvalues $\left(  2\lambda^{2},\lambda^{2}\right)  $.
This point repels nearby orbits lying on the cylinder for all $\lambda\neq0$.
However, if we take into account the stability along $r$, the point is
generically a saddle.
\end{itemize}

\section{Asymptotic expansions in the regime where the cubic derivative term
dominates}

\label{asym}

An important question about the present model is: in what phase of the
cosmological evolution are the extra cubic interaction terms expected to play
an important role? Another interesting question is: how does the cosmological
predictions of our model differ from those of standard FRW cosmology driven by
a scalar field with exponential potential? To answer these questions we
proceed as follows.

To show in what phase of the cosmological history it is expected that the
cubic derivatives play a role, let us consider $g_{0}\gg1$, and take the limit
$g_{0}$ to infinity in the equations for $\dot{H},\ddot{\phi}$. We obtain the
approximations
\begin{equation}
\ddot{\phi}+\frac{1}{2}\lambda\dot{\phi}^{2}=0,\quad\dot{H}-\lambda H\dot
{\phi}+3H^{2}+\frac{1}{6}\lambda^{2}\dot{\phi}^{2}=0
\end{equation}
that admits the solution
\begin{equation}
H_{s}(t)=\frac{6H_{0}-\lambda\dot{\phi}_{0}}{3\left(  (t-t_{0})\left(
6H_{0}-\lambda\dot{\phi}_{0}\right)  +2\right)  }+\frac{\lambda\dot{\phi}_{0}%
}{3\lambda\dot{\phi}_{0}(t-t_{0})+6},\quad\phi_{s}(t)=\phi_{0}+\frac
{2}{\lambda}\ln\left(  1+\frac{\lambda\dot{\phi}_{0}}{2}(t-t_{0})\right)  ,
\end{equation}
where we have chosen the initial conditions $H(t_{0})=H_{0},\;\phi(t_{0}%
)=\phi_{0},\;\dot{\phi}(t_{0})=\dot{\phi}_{0}$.\newline Furthermore, we have
\begin{equation}
a_{s}(t)=a_{0}\left(  1+\frac{1}{2}\lambda\dot{\phi}_{0}(t-t_{0})\right)
^{\frac{1}{3}}\left(  1+\frac{1}{2}(t-t_{0})\left(  6H_{0}-\lambda\dot{\phi
}_{0}\right)  \right)  ^{\frac{1}{3}}.
\end{equation}
Taking initial conditions such that $\dot{\phi}_{0}=0$ or $6H_{0}-\lambda
\dot{\phi}_{0}=0$, we obtain that $a(t)\propto t^{\frac{1}{3}}$, that is, it
corresponds to a radiation dominated universe. On the other hand, if we choose
initial conditions such that $3H_{0}-\lambda\dot{\phi}_{0}=0$, we obtain that
$a(t)\propto t^{\frac{2}{3}}$, that is, it corresponds to a universe dominated
by dark matter. For other values of the initial conditions we can model the
transition from a radiation dominated universe to a matter dominated one.
However, from the leading terms in the Friedmann equations as $g_{0}%
\rightarrow\infty$ we obtain
\begin{equation}
-\frac{1}{2}a_{0}^{3}\dot{\phi}_{0}^{3}e^{\lambda\phi_{0}}\left(  \lambda
\dot{\phi}_{0}-6H_{0}\right)  =0.
\end{equation}
Since we are looking for universes with $a_{0}>0$, it immediately follows that
Galileon modifications are particularly relevant for the radiation-dominated
(early time universe) as shown in \cite{genlyGL}. On the other hand, for
$g_{0}\rightarrow0$, the model is well-suited for describing the late-time
universe, and we recover the standard quintessence results found elsewhere,
e.g., in the seminal work \cite{Copeland:1997et}. Now let us use the above
solution as a seed solution to construct an asymptotic expansion for large
$g_{0}$ when the potential is turned on.

Let us define
\begin{equation}
H=H_{s}+\varepsilon h+\mathcal{O}(\varepsilon^{2})\quad\phi=\phi
_{s}+\varepsilon\Phi+\mathcal{O}(\varepsilon^{2}),\quad\varepsilon=g_{0}^{-1}.
\end{equation}
and let us assume $\dot{\phi}_{0}\neq0$ (without losing generality we can set
$\phi_{0}=0$). Then we have at zeroth-order the equation
\begin{equation}
\frac{4\dot{\phi}_{0}^{3}\left(  \lambda\dot{\phi}_{0}-6H_{0}\right)
}{\left(  \lambda\dot{\phi}_{0}(t-t_{0})+2\right)  \left(  -6H_{0}%
(t-t_{0})+\lambda\dot{\phi}_{0}(t-t_{0})-2\right)  }=0.
\end{equation}
Since we have assumed $\dot{\phi}_{0}\neq0$, the solution is $H_{0}%
=\frac{\lambda}{6}\dot{\phi}_{0}$. This implies that $a(t)\propto t^{\frac
{2}{3}}$. That is, it corresponds to a universe dominated by a dust fluid/dark
matter. Substituting back the value for $H_{0}$ in the Klein-Gordon, the
Raychaudhuri and Friedmann equations, respectively, we have the following
first-order equations:
\begin{subequations}
\label{eqxx21}%
\begin{align}
\dot{\phi}_{0}^{2}\left(  -\lambda^{2}+3\lambda\dot{\phi}_{0}\left(
(t-t_{0})\Phi^{\prime\prime}(t)\left(  \lambda\dot{\phi}_{0}(t-t_{0}%
)+4\right)  +2\Phi^{\prime}(t)\left(  \lambda\dot{\phi}_{0}(t-t_{0})+2\right)
\right)  +12\Phi^{\prime\prime}(t)+6\right)  -12V_{0}  &  =0,\\
9h^{\prime}(t)\left(  \lambda\dot{\phi}_{0}(t-t_{0})+2\right)  {}^{2}%
+\lambda\left(  -\lambda^{2}+3\lambda\dot{\phi}_{0}\Phi^{\prime}(t)\left(
\lambda\dot{\phi}_{0}(t-t_{0})+2\right)  +6\right)   &  =0,\\
2\left(  \dot{\phi}_{0}^{2}\left(  -3\dot{\phi}_{0}\left(  \lambda\dot{\phi
}_{0}(t-t_{0})+2\right)  \left(  \lambda\Phi^{\prime}(t)-6h(t)\right)
+\lambda^{2}-6\right)  -12V_{0}\right)   &  =0.
\end{align}

The solution of \eqref{eqxx21} that satisfies $\Phi(t_{0})=\Phi_{0}%
,h(t_{0})=h_{0}$ is given by:
\end{subequations}
\begin{subequations}
\begin{align}
&  \Phi=\Phi_{0}+\frac{4\left(  2V_{0}-3h_{0}\dot{\phi}_{0}^{3}\right)
}{\lambda^{2}\dot{\phi}_{0}^{4}\left(  \frac{1}{2}\lambda\dot{\phi}%
_{0}(t-t_{0})+1\right)  }-\frac{4\left(  2V_{0}-3h_{0}\dot{\phi}_{0}%
^{3}\right)  }{\lambda^{2}\dot{\phi}_{0}^{4}}+\frac{\left(  \left(
\lambda^{2}-6\right)  \dot{\phi}_{0}^{2}+12V_{0}\right)  \ln\left(  \frac
{1}{2}\lambda\dot{\phi}_{0}(t-t_{0})+1\right)  }{3\lambda^{2}\dot{\phi}%
_{0}^{4}},\\
&  h=\frac{3h_{0}\dot{\phi}_{0}^{3}-2V_{0}}{3\dot{\phi}_{0}^{3}\left(
\frac{1}{2}\lambda\dot{\phi}_{0}(t-t_{0})+1\right)  {}^{2}}+\frac{2V_{0}%
}{3\dot{\phi}_{0}^{3}\left(  \frac{1}{2}\lambda\dot{\phi}_{0}(t-t_{0}%
)+1\right)  },
\end{align}
where we observe that$~\lim_{t\rightarrow\infty}\Phi(t)=\infty\;$%
unless$\;\frac{V_{0}}{\dot{\phi}_{0}^{2}}=\frac{6-\lambda^{2}}{12}$,
and~$\lim_{t\rightarrow\infty}h(t)=0.$ Furthermore\ the relative errors are
defined to be%

\end{subequations}
\begin{equation}
E(H_{s})=\frac{|H-H_{s}|}{H},\quad E(\phi_{s})=\frac{|\phi-\phi_{s}|}{\phi
},\quad E(\dot{H}_{s})=\frac{|\dot{H}-\dot{H}_{s}|}{\dot{H}},\quad E(\dot
{\phi}_{s})=\frac{|\dot{\phi}-\dot{\phi}_{s}|}{\dot{\phi}},
\end{equation}
and satisfies the conditions
\begin{subequations}
\begin{equation}
\lim_{t\rightarrow\infty}E(\Phi_{s}(t))=\lim_{t\rightarrow\infty}E(\dot{\Phi
}_{s}(t))=\frac{\epsilon\left(  \left(  \lambda^{2}-6\right)  \dot{\phi}%
_{0}^{2}+12V_{0}\right)  }{6\lambda\dot{\phi}_{0}^{4}},
\end{equation}%
\end{subequations}
\[
\lim_{t\rightarrow\infty}E(H_{s}(t))=\lim_{t\rightarrow\infty}E(H_{s}%
(t))=\frac{2V_{0}\epsilon}{\lambda\dot{\phi}_{0}^{4}}.
\]
Thus, the relative errors can be small enough for $V_{0}\ll\dot{\phi}_{0}%
^{4},\lambda=\pm\sqrt{6}$ or $V_{0}\ll\dot{\phi}_{0}^{4},\dot{\phi}_{0}^{2}%
\gg1$ (the last condition is fulfilled for all finite values of $V_{0}$ and
large values of $\dot{\phi}_{0}$). We see that for $V_{0}\gg\dot{\phi}_{0}%
^{4}$ and finite $\dot{\phi}_{0}$, the relative errors are large, and the
approximation fails. In the former case the potential makes the ideal gas
solution unstable, as expected for a universe dominated by a dust fluid/dark
matter. \ That is an important result since the cubic term in the Galileon
field provides a matter epoch.

\section{The Galileon model with matter}

\label{matter}

Until now we have considered the case with vacuum (i.e. $\rho_{m}=0$), and we have found at some fixed points
$c_{s}^{2}<0$, and $Q_{S}<0$ also. On the other hand, we require for the
avoidance of Laplacian instabilities $c_{s}^{2}\geq0$ and for the absence of
ghosts it is required that $Q_{S}>0$. In this section we study if the matter
stabilizes the Galileon field, in the sense that it helps to restore the above
conditions. To begin with, the definition of $c_{s}^{2}$ changes to
$~c_{S}^{2}\equiv\frac{6w_{1}H-3w_{1}^{2}-6\dot{w}_{1}-6\rho_{m}}%
{4w_{2}+9w_{1}^{2}};~$when $\rho_{m}>0$ and $w_{tot}\neq w_{DE}$.

By using the variables
\begin{equation}
x=\frac{\dot{\phi}}{\sqrt{6(H^{2}+1)}},y=\frac{V_{0}e^{-\lambda\phi}}%
{3(H^{2}+1)},z=\frac{H}{\sqrt{H^{2}+1}},
\end{equation}
and the parameter $\alpha=g_{0}V_{0}$, we obtain a dynamical system the same
as before \eqref{eq.38}:
\begin{equation}
x^{\prime}=\frac{f_{1}(\mathbf{x})}{f_{4}(\mathbf{x})},~y^{\prime}=\frac
{f_{2}(\mathbf{x})}{f_{4}(\mathbf{x})},~z^{\prime}=\frac{f_{3}(\mathbf{x}%
)}{f_{4}(\mathbf{x)}},
\end{equation}
$\mathbf{x=}\left(  x,y,z\right)  $ and functions $f_{1}-f_{4}$ are\thinspace
\ defined by \eqref{eq.38b}. We have an auxiliary evolution equation for
\begin{equation}
\tilde{\Omega}_{m}=\frac{\rho_{m}}{3(H^{2}+1)},\quad\rho_{m}=\rho_{m,0}a^{-3}.
\end{equation}
The system can be written in the compact form
\begin{subequations}
\begin{align}
&  x^{\prime}=\frac{2qz^{2}\left(  6\alpha x^{3}z+\sqrt{6}y\right)  -6\sqrt
{6}\alpha\lambda x^{4}+12\alpha x^{3}z^{3}-3\sqrt{6}x^{2}y+\sqrt{6}y\left(
3y-z^{2}\right)  }{12\alpha x^{2}},\\
&  y^{\prime}=y\left(  2(q+1)z^{3}-\sqrt{6}\lambda x\right)  ,\\
&  z^{\prime2}=\left(  z^{2}-1\right)  ,\\
&  \tilde{\Omega}_{m}^{\prime}=\tilde{\Omega}_{m}z\left(  2(q+1)z^{2}%
-3\right)
\end{align} where $q$, the deceleration parameter, can be expressed in terms of the phase-space variables $(x,y,z)$.
Observe that the evolution equation for matter decouples from the other
evolution equations. The subtle difference, including dust matter, is that the
restriction \eqref{restr_1} now becomes%

\end{subequations}
\begin{equation}
y^{2}+2x^{3}\alpha(-\sqrt{6}z+x\lambda)+y(x^{2}-z^{2}+\tilde{\Omega}_{m})=0.
\end{equation}

Since $y\geq0,\tilde{\Omega}_{m}\geq0$, the motion of the particle in the
phase space is not just restricted to the given surface $y^{2}+2x^{3}%
\alpha(-\sqrt{6}z+x\lambda)+y(x^{2}-z^{2})=0$, as it is the case in
\eqref{eq.38} and \eqref{eq.38b}. Instead, the phase space is now the set
\begin{equation}
\left\{  (x,y,z):y^{2}+2x^{3}\alpha(-\sqrt{6}z+x\lambda)+y(x^{2}-z^{2}%
)\leq0,y\geq0\right\}  .
\end{equation}
There are recovered the fixed points for the vacuum case, that is, with
$\tilde{\Omega}_{m}=0;$ but some existence conditions change which is the main
difference between the two cases. The existence conditions for $P_{10},P_{11}$
changes to $\alpha/\lambda\geq0$; the existence conditions of $P_{12},P_{13}$
are changed by $\lambda^{2}+3\alpha\lambda-3\geq0$. The stability conditions
changes accordingly. The eigenvalues and stability conditions for
$P_{10},P_{11}$ remains unaffected when matter is included. The eigenvalues
for $P_{12,13}$ changes to $\pm\left(  -3,\frac{3}{4}-\frac{3\sqrt
{\lambda(6\alpha(\alpha\lambda-4)-7\lambda)+24}}{4\sqrt{6\alpha^{2}+1}\lambda
},\frac{3}{4}+\frac{3\sqrt{\lambda(6\alpha(\alpha\lambda-4)-7\lambda)+24}%
}{4\sqrt{6\alpha^{2}+1}\lambda}\right)  $, but the point will still be a
saddle. The results shown in table \ref{tab:Table_II} remains unaltered. We
conclude then, that the matter background has no imprint in the Galileon field
concerning the avoidance of Laplacian instabilities and concerning the absence
of ghosts. The conditions $c_{s}^{2}\geq0,Q_{S}>0$ are not restored for the
points that violate them.\newline

\section{Conclusions}

\label{conc}

In this paper we have studied cubic Galileon cosmology with an extra
conservation law from the dynamical systems perspective. The novelty of this
model is that was derived by the application of Noether's theorem in the
gravitational Lagrangian. Firstly, we have noticed that the fixed points
$A^{\pm},B^{\pm},C$ and $D$ investigated in detail in \cite{genlyGL}, do not
exist in our scenario. So, our analysis has its own right, and it is
complementary to all the analysis done before. This new scenario admits
power-law solutions. We have found a new asymptotic solution given by the
fixed point solution $P_{1}$, which satisfies
\[
H\approx\frac{3}{8}b_{1}^{2}\left(  -\frac{3\ln t}{t^{2}}-\frac{4}{b_{2}%
-b_{1}t}\right) ,\quad\phi\approx\frac{\ln\left( \frac{4 V_{0}}{4
\alpha\lambda^{3} (2 a_{1}-3 b_{1})^{2}+2 \lambda^{2} (2 a_{1}-3 b_{1}%
)^{2}+3b_{1} (9 b_{1}-4)}\right) }{\lambda}+\frac{2 \ln t}{\lambda},
\]
while the scale factor is calculated to be $a(t)\approx a_{0}e^{\frac
{9b_{1}^{2}}{8t}}t^{\frac{9b_{1}^{2}}{8t}}(b_{2}-b_{1}t)^{3b_{1}/2}
=(b_{2}-b_{1}t)^{3b_{1}/2}\left(  a_{0}+\mathcal{O}\left(  t^{-1})\right)
\right) $.

This point has not been obtained previously in \cite{genlyGL} and in
\cite{genlyGL2}, since in these works the authors used $H$-normalization,
which fails obviously when $H=0$. This solution attracts the exact power-law
solution previously described in \cite{Dimakis:2017kwx} for the proper choice
of free parameters. Moreover, we have obtained several solutions ($P_{1}%
$-$P_{5}$, $P_{8},P_{9}$, $P_{14}(x_{c})$, $P_{16}(z_{c})$), that violates the
conditions $c_{s}^{2}\geq0,Q_{S}>0$ (one or both) suffering from Laplacian
instabilities and from the presence of ghosts, that makes them physically less
interesting. Excluding the de Sitter solutions, all of them mimic dust
solutions. However, we have found other solutions that are free of
Laplacian instabilities, and are ghosts-free. For instance:

\begin{itemize}
\item The solutions $P_{6,7}(x_{c})$ can satisfy the conditions $c_{s}^{2}%
\geq0, Q_{S}>0$, denoted by the yellow region with dot-dashed boundary in the
Figure \ref{fig:RegionP6}. Although,
they can violate these conditions, for example, it is possible to have
$c_{s}^{2}< 0, Q_{S}\leq0$ as represented by the
green region surrounded by a solid line in the Figure \ref{fig:RegionP6}.

\item The fine-tuned solutions $P_{10}$ and $P_{11}$, that exists in the
unmodified (quintessence) scenario can be sink, respectively, source for
$0\leq\lambda^{2}\leq\frac{3}{2}$, and they satisfy $c_{s}^{2}\geq0,Q_{S}>0$
for $0<\lambda^{2}<2$.

\item The solutions $P_{12}$ and $P_{13}$ that exists for $\lambda^{2}%
+3\alpha\lambda-3=0$, mimics dust and they are saddles. We have proved that
$c_{s}^{2}\geq0,Q_{S}>0$ for $1\leq\lambda^{2}<\frac{3}{11}\left(  3+\sqrt
{53}\right)  \approx2.80367$. In this region of the parameter space, the
cosmological solution is free of Laplacian instabilities and ghosts.

\item The line of fixed points $P_{15}(z_{c}):\left(  \beta z_{c},\sqrt
{6}\alpha\beta{z_{c}}^{2},z_{c}\right)  $ exists for $\lambda=0, \beta
=\frac{1}{\sqrt{2}}\left(  \sqrt{3}\alpha-\sqrt{3\alpha^{2}-2}\right) $, and
the cosmological solution is given by
\[
\dot\phi(t)\approx\frac{\sqrt{6} \beta{z_{c}}}{\sqrt{1-{z_{c}}^{2}}}, \quad
H(t)\approx\frac{{z_{c}}}{\sqrt{1-{z_{c}}^{2}}}, \quad\phi(t)-\phi_{0}%
\approx\frac{\sqrt{6} \beta{z_{c}} t}{\sqrt{1-{z_{c}}^{2}}}, \quad a(t)\approx
a_{0} e^{\frac{t z_{c}}{\sqrt{1-{z_{c}}^{2}}}}, \quad z_{c}=\pm\frac
{\sqrt{V_{0}}}{\sqrt{3 \sqrt{6} \alpha\beta+V_{0}}}.
\]
This solution is free of Laplacian instabilities and ghosts-free in the region
displayed in figure \ref{fig:RegionP15}.
\end{itemize}

Furthermore we have investigated the fixed points at infinity, and all of them
are saddle points, so they have no relevance for the early or late time universe.

The Galileon modifications are particularly relevant for the
radiation-dominated (early time universe) as shown in \cite{genlyGL} by
investigating the regime where the coupling parameter satisfy $g_{0}\gg1$. On
the other hand, for $g_{0}\rightarrow0$ the model is well-suited for
describing the late-time universe, and we have recovered the standard
quintessence results found elsewhere, e.g., in the seminal work
\cite{Copeland:1997et}. We have constructed asymptotic expansions of the
power-law solution for large $g_{0}$ when the potential is turned on.

All the previous results were found for vacuum Galileon ($\rho_{m}=0$). We
have seen that for some points $c_{s}^{2}<0$, and at some fixed points
$Q_{S}<0$ also. Since we required to avoid Laplacian instabilities we must
have $c_{s}^{2}\geq0$ and if we demand the absence of ghosts it is required
that $Q_{S}>0$. Moreover, we have investigated if the matter stabilizes the Galileon
field, in the sense that it helps to restore the above conditions. When we
include background matter in the form of dust, we recovered the fixed points
for $\tilde{\Omega}_{m}=0$ but some existence conditions change, especially
for $P_{10},P_{11}$ and $P_{12},P_{13}$. The stability conditions changes
accordingly. The eigenvalues and stability conditions for $P_{10},P_{11}$
remains unaffected when matter is included. The eigenvalues for $P_{12,13}$
changes, but the point will still be a saddle. The results shown in table
\ref{tab:Table_II} remains unaltered. We conclude then, that the matter
background has no imprint in the Galileon field concerning the avoidance of
Laplacian instabilities and concerning the absence of ghosts. The conditions
$c_{s}^{2}\geq0,Q_{S}>0$ are not restored for the points that violate them.

\begin{figure}
	\centering
		\includegraphics[width=0.7\textwidth]{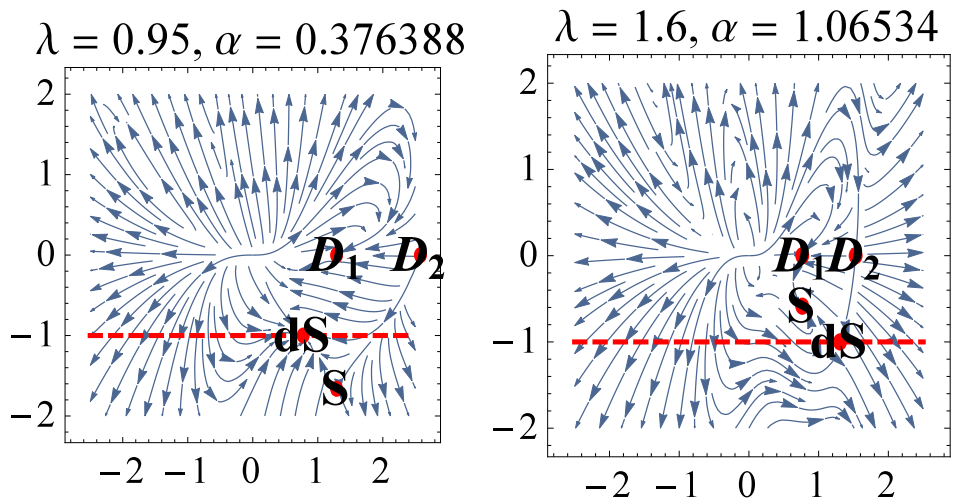}
	\caption{\label{fig:wtot_vs_x} Streamlines in the plane $(w_{tot},x)$, representing the typical behavior of our model for $\alpha=\frac{3}{4 \lambda  \left(3-\lambda ^2\right)}$ and different choices of $\lambda$.}
	\end{figure}

{Finally, in order to get a cosmologically suitable scenario we require  $z>0$ (i.e., $H>0$) for accommodating a late-time accelerated expansion phase.
For simplicity, we will restrict the analysis to the invariant set of the flow of the system \eqref{eq.38} given by $z=1$.
In this invariant set the total equation of state parameter is $w_{tot}= -y$. Now, for the special parameters choice $\alpha=\frac{3}{4 \lambda  \left(3-\lambda ^2\right)}$, the fixed points $P_7(x_c)$ merges to one fixed point, denoted by $dS$, in the plane $(x,w_{tot})$ with coordinates $(x,w_{tot})=\left(\sqrt{\frac{2}{3}} \lambda , -1\right)$ that mimics a de Sitter solution. It is an attractor for $0<\lambda <\sqrt{\frac{3}{2}}$. For $0.72262<\lambda <1.22474$ the corresponding cosmological solution satisfies the physical conditions $c_s^2> 0, Q_S\geq 0$.
We obtain also a transient phase given by the fixed point $D_1:= (x,w_{tot})=\left(\frac{\sqrt{\frac{3}{2}}}{\lambda }, 0\right)$, that mimics a dust-like solution. Furthermore, we have a scaling solution $S:=(x,w_{tot})=\left(\frac{\sqrt{\frac{3}{2}}}{\lambda }, -\frac{3}{2 \lambda ^2}\right)$, which can be an attractor for $\lambda <-\frac{3}{\sqrt{2}}$  or $\sqrt{\frac{3}{2}}<\lambda <\sqrt{3}$, or a saddle otherwise. The scaling solution is accelerating for $-\frac{3}{\sqrt{2}}<\lambda <-\sqrt{3}$ or $0<\lambda <\sqrt{3}$. The physical conditions $c_s^2>0,Q_S\geq 0$ are satisfied for $0.662827<\lambda <1.60021$. The scaling solution and the de Sitter are not attractors simultaneously. Besides, the model admits an additional dust-like solution $D_2:= (x, w_{tot})=\left(\frac{\sqrt{6}}{\lambda },0\right)$ which is unstable.
In the Figure \ref{fig:wtot_vs_x} it is portrayed in the plane $w_{tot}$ vs. $x$, the typical behavior of our model for $\alpha=\frac{3}{4 \lambda  \left(3-\lambda ^2\right)}$ and different choices of $\lambda$. On the left panel the attractor of the de Sitter point, $dS$, such that the value $w_{tot}=-1$ is approached asymptotically. However, we immediately see that there are trajectories connecting the dust point $D_2$ and the scaling solution $S$ that crosses the phantom divide line (represent by a red dashed line), that eventually enter the $w=-1$ region. However, some solutions are trapped on the phantom regime. On the right panel, the scaling point $S$ is a stable spiral, and the de Sitter point is a transient one. There are trajectories connecting the dust point $D_2$ and the de Sitter solution $dS$, crossing the phantom divide. As in the previous case, some orbits remains on the phantom region.  As we have briefly shown, this model has quite interesting cosmological features, resembling the usual late-time dynamics usually found in conventional quintessence models for a non-zero value of $\alpha$.}

Different subjects of study are still open for the integrable model in which
new critical points exist. Hence, the requirement of the existence of a
conservation law has a physical interpretation in the evolution of the model.
However the exact nature of the physical interpretation for this model is
still unknown, such an analysis is still in progress.

\begin{acknowledgments}
This work was financial supported by FONDECYT grants, 1150246 (AG) and 3160121
(AP). \ AP thanks the Durban University of Technology and the University of
the Witwatersrand for the hospitality provided while this work was performed.
SJ would like to acknowledge the financial support from the National Research
Foundation of South Africa with grant number 99279. AP and SJ also acknowledge
support from the DST-NRF Centre of Excellence in Mathematical and Statistical
Sciences (CoE-MaSS). Opinions expressed and conclusions arrived at are those
of the authors and are not necessarily to be attributed to the CoE-MaSS. GL
acknowledges DI-VRIEA for financial support through Proyectos VRIEA
Investigador Joven 2016 and Investigador Joven 2017.
\end{acknowledgments}

\appendix

\section{Nonhyperbolic fixed point at infinity}

\label{analysisQ1} For analyzing the stability of $Q_{1}$ we proceed as
follows. Dividing by the positive quantity $\lambda^{2}\cos^{2}(\theta)$, we
have that the flow of the system \eqref{eq17} is equivalent to the flow of
\begin{subequations}
\begin{align}
&  \frac{d\theta}{d T}=\hat{h}_{1}(\theta,z):= z \sin(\theta) \cos(\theta),\\
&  \frac{dz}{d T}=\hat{h}_{2}(\theta,z):= \left(  z^{2}-1\right)  ,\\
&  \frac{dr}{d T}=\hat{h}_{3}(r,\theta,z):=r\left[  z (\cos(2 \theta
)-3)\right]  .
\end{align}
To improve the range and accuracy we calculate the diagonal first order Pade
approximants
\end{subequations}
\[
[1/1]_{\hat{h}_{1}}(\theta),\quad[1/1]_{\hat{h}_{2}}(\theta),\quad
[1/1]_{\hat{h}_{3}}(\theta),
\]
around $\theta=\pi/2$, which yields the following approximate expressions
\begin{equation}
\frac{d\theta}{d T}\approx\frac{1}{2} (\pi-2 \theta) z,\quad\frac{d z}{d
T}\approx-1+z^{2},\quad\frac{dr}{d T}\approx-4 r z.
\end{equation}
Imposing the initial conditions
\[
z(0)=-1+\delta z, \quad\theta(0)=\frac{\pi}{2}+\delta\theta, \quad r(0)=\delta
r, \quad\delta z\ll1, \delta\theta\ll1, \delta r\ll 1,
\]
we obtain the solution
\begin{equation}
z=\delta z e^{-2 T }-1,\quad\theta=\delta\theta e^{T }+\frac{\pi}{2},
\quad r=\delta r e^{4 T}.
\end{equation}
Observe that the small perturbations $\delta\theta,$ and $\delta r$ grow
exponentially, and $z\rightarrow-1$ as $T\rightarrow+\infty$. At the cylinder
at infinity ($\delta r=0$) the orbits near $Q_{1}$ are attracted by
the invariant set $z=-1$, but the angle departs from the equilibrium value
$\theta=\frac{\pi}{2}$ as time goes forward. This means that $Q_{1}$ is the
local past attractor at the circle $z=1, X^{2}+Y^{2}=1, Y\geq0$, as it is
shown in figure \ref{fig:ProblemCFig1}. If we move to the interior of the
phase space $\delta r>0$, the orbits are repelled by the cylinder at infinity
at an exponential rate, although $z\rightarrow-1$ as $T\rightarrow+\infty$.
Hence, $Q_{1}$ is generically a saddle.

\section{Fixed points lying on singular surfaces}

\label{appendixA} The system \eqref{eq.38} blows-up when $f_{4}(x,y,z)=0$, and
specially at the fixed points $P_{1}, P_{8}, P_{9}$. For this reason we had
examined their stability numerically or by taking limits, since they lead to
indeterminacy of the form $0/0$.

For examining the stability of $P_{1}$ we introduce the new variables $u=
z-\frac{\sqrt{\frac{3}{2}} x}{\lambda}, \quad v= \frac{2 z}{3}, \quad w= y$,
such that the system can be written in the canonical form
\begin{align}
\label{P1secondorder} &  \left(
\begin{array}
[c]{c}%
u^{\prime}\\
v^{\prime}\\
w^{\prime}\\
\end{array}
\right)  = \left(
\begin{array}
[c]{ccc}%
0 & 0 & 0\\
0 & 0 & 1\\
0 & 0 & 0\\
&  &
\end{array}
\right)  \left(
\begin{array}
[c]{c}%
u\\
v\\
w\\
\end{array}
\right) \nonumber\\
&  + \left(
\begin{array}
[c]{c}%
\lambda u^{2} \left(  \alpha\left(  2 \lambda^{2}-3\right)  -\lambda\right)
+\frac{3}{2} u v \left(  2 \lambda\left(  \lambda-2 \alpha\left(  \lambda
^{2}-3\right)  \right)  -3\right)  +\frac{9}{8} v^{2} \left(  2 \alpha\left(
2 \lambda^{2}-9\right)  \lambda-2 \lambda^{2}+3\right) \\
\frac{1}{12} \left(  -4 \alpha\lambda^{3} (2 u-3 v)^{2}-2 \lambda^{2} (2 u-3
v)^{2}-27 v^{2}\right) \\
\lambda^{2} w (2 u-3 v)
\end{array}
\right)  +\mathcal{O}(|(u,v,w)^{T}|^{3}).
\end{align}

%

Given $H^{2}$, the vector space of the homogeneous polynomials of second order
in $\mathbf{u}=(u_{1},u_{2},u_{3})$, let us consider the linear operator
$\mathbf{L_{J}^{(2)}}$ associated to $\mathbf{J}=\left(
\begin{array}
[c]{ccc}%
0 & 0 & 0\\
0 & 0 & 1\\
0 & 0 & 0
\end{array}
\right) $, that assigns to $\mathbf{h(u)}\in H^{2}$, the Lie bracket of the
vector fields $\mathbf{Ju}$ and $\mathbf{h(u)}$:
\begin{align}
\mathbf{L_{J}^{(2)}}:H^{2}  &  \rightarrow H^{2}\nonumber\\
\mathbf{h}  &  \rightarrow\mathbf{L_{J}}\mathbf{h(u)}=\mathbf{Dh(u)}%
\mathbf{Ju}-\mathbf{Jh(u)}.
\end{align}
where $H^{2}$ the real vector space of vector fields whose components are
homogeneous polynomials of degree 2. The canonical basis for the real vector
space of 3-dimensional vector fields whose components are homogeneous
polynomials of degree 2 is given by
\begin{align}
&  H^{2}=\text{span}\left\{  \left(
\begin{array}
[c]{c}%
u_{1}^{2}\\
0\\
0
\end{array}
\right)  ,\,\left(
\begin{array}
[c]{c}%
u_{1}u_{2}\\
0\\
0
\end{array}
\right)  ,\,\left(
\begin{array}
[c]{c}%
u_{1}u_{3}\\
0\\
0
\end{array}
\right)  ,\,\left(
\begin{array}
[c]{c}%
u_{2}^{2}\\
0\\
0
\end{array}
\right)  ,\,\left(
\begin{array}
[c]{c}%
u_{2}u_{3}\\
0\\
0
\end{array}
\right)  ,\,\left(
\begin{array}
[c]{c}%
u_{3}^{2}\\
0\\
0
\end{array}
\right)  ,\right. \nonumber\\
&  \left.  \quad\quad\quad\ \ \ \ \ \left(
\begin{array}
[c]{c}%
0\\
u_{1}^{2}\\
0
\end{array}
\right)  ,\,\left(
\begin{array}
[c]{c}%
0\\
u_{1}u_{2}\\
0
\end{array}
\right)  ,\,\left(
\begin{array}
[c]{c}%
0\\
u_{1}u_{3}\\
0
\end{array}
\right)  ,\,\left(
\begin{array}
[c]{c}%
0\\
u_{2}^{2}\\
0
\end{array}
\right)  ,\,\left(
\begin{array}
[c]{c}%
0\\
u_{2}u_{3}\\
0
\end{array}
\right)  ,\,\left(
\begin{array}
[c]{c}%
0\\
u_{3}^{2}\\
0
\end{array}
\right)  \right. \nonumber\\
&  \left.  \quad\quad\quad\ \ \ \ \ \left(
\begin{array}
[c]{c}%
0\\
0\\
u_{1}^{2}%
\end{array}
\right)  ,\,\left(
\begin{array}
[c]{c}%
0\\
0\\
u_{1}u_{2}%
\end{array}
\right)  ,\,\left(
\begin{array}
[c]{c}%
0\\
0\\
u_{1}u_{3}%
\end{array}
\right)  ,\,\left(
\begin{array}
[c]{c}%
0\\
0\\
u_{2}^{2}%
\end{array}
\right)  ,\,\left(
\begin{array}
[c]{c}%
0\\
0\\
u_{2}u_{3}%
\end{array}
\right)  ,\,\left(
\begin{array}
[c]{c}%
0\\
0\\
u_{3}^{2}%
\end{array}
\right)  \right\}  .
\end{align}

By computing the action of $\mathbf{L_{J}^{(2)}}$ on each basis element on
$H^{2}$ we have%

\begin{align}
&  \mathbf{L_{J}^{(2)}}\left(  H^{2}\right)  =\text{span}\left\{  \left(
\begin{array}
[c]{c}%
u_{1}u_{3}\\
0\\
0
\end{array}
\right)  ,\,\left(
\begin{array}
[c]{c}%
u_{2}u_{3}\\
0\\
0
\end{array}
\right)  ,\,\left(
\begin{array}
[c]{c}%
u_{3}^{2}\\
0\\
0
\end{array}
\right)  ,\,\left(
\begin{array}
[c]{c}%
0\\
u_{1}^{2}\\
0
\end{array}
\right)  ,\,\left(
\begin{array}
[c]{c}%
0\\
u_{1}u_{2}\\
0
\end{array}
\right)  ,\,\left(
\begin{array}
[c]{c}%
0\\
u_{1}u_{3}\\
0
\end{array}
\right)  ,\right. \nonumber\\
&  \left.  \quad\quad\quad\quad\quad\quad\ \ \ \ \ \left(
\begin{array}
[c]{c}%
0\\
u_{2}^{2}\\
0
\end{array}
\right)  ,\,\left(
\begin{array}
[c]{c}%
0\\
u_{2}u_{3}\\
0
\end{array}
\right)  ,\,\left(
\begin{array}
[c]{c}%
0\\
u_{3}^{2}\\
0
\end{array}
\right)  ,\,\left(
\begin{array}
[c]{c}%
0\\
0\\
u_{1}u_{3}%
\end{array}
\right)  ,\,\left(
\begin{array}
[c]{c}%
0\\
0\\
u_{2}u_{3}%
\end{array}
\right)  ,\,\left(
\begin{array}
[c]{c}%
0\\
0\\
u_{3}^{2}%
\end{array}
\right)  \right\}  . \label{nonresonant2}%
\end{align}
Thus, the second order terms that are linear combinations of the six vectors
in \eqref{nonresonant2} can be eliminated \cite{wiggins}. To determine the
nature of the second order terms that cannot be eliminated we must compute the
complementary space of \eqref{nonresonant2} which is
\[
G^{2}=\text{span}\left\{  \left(
\begin{array}
[c]{c}%
u_{1}^{2}\\
0\\
0
\end{array}
\right)  ,\,\left(
\begin{array}
[c]{c}%
u_{1}u_{2}\\
0\\
0
\end{array}
\right)  ,\,\left(
\begin{array}
[c]{c}%
u_{2}^{2}\\
0\\
0
\end{array}
\right)  ,\,\left(
\begin{array}
[c]{c}%
0\\
0\\
u_{1}^{2}%
\end{array}
\right)  ,\,\left(
\begin{array}
[c]{c}%
0\\
0\\
u_{1}u_{2}%
\end{array}
\right)  ,\,\left(
\begin{array}
[c]{c}%
0\\
0\\
u_{2}^{2}%
\end{array}
\right)  \right\}  .
\]
Following the above reasoning, we propose a transformation
\begin{equation}
\left(
\begin{array}
[c]{c}%
u\\
v\\
w
\end{array}
\right)  =\left(
\begin{array}
[c]{c}%
\sigma_{1}\\
\sigma_{2}\\
\sigma_{3}%
\end{array}
\right)  +\left(
\begin{array}
[c]{c}%
h_{2}^{(1)}(\sigma_{1},\sigma_{2},\sigma_{3})\\
h_{2}^{(2)}(\sigma_{1},\sigma_{2},\sigma_{3})\\
h_{2}^{(3)}(\sigma_{1},\sigma_{2},\sigma_{3})
\end{array}
\right)
\end{equation}
where the $h_{2}^{(i)},\;{i=1,2,3}$ are homogeneous polynomials of degree 2 in
$(\sigma_{1},\sigma_{2},\sigma_{3})$. Up to this point $h_{2}^{(i)}%
,\;{i=1,2,3}$ are completely arbitrary. Now we choose a specific form of them
so as to simplify the $\mathcal{O}(2)$ terms as much as possible. We choose
\begin{equation}
\left(
\begin{array}
[c]{c}%
h_{2}^{(1)}(\sigma_{1},\sigma_{2},\sigma_{3})\\
h_{2}^{(2)}(\sigma_{1},\sigma_{2},\sigma_{3})\\
h_{2}^{(3)}(\sigma_{1},\sigma_{2},\sigma_{3})
\end{array}
\right)  =\left(
\begin{array}
[c]{c}%
0\\
0\\
\frac{2}{3}\lambda^{2}\sigma_{1}^{2}(2\alpha\lambda+1)-2\lambda^{2}\sigma
_{1}\sigma_{2}(2\alpha\lambda+1)+\frac{3}{4}\sigma_{2}^{2}\left(
4\alpha\lambda^{3}+2\lambda^{2}+3\right) \\
\end{array}
\right)  .
\end{equation}
Hence, the normal form of the system \eqref{P1secondorder} is
\begin{subequations}
\begin{align}
&  \sigma_{1}^{\prime}=\lambda\sigma_{1}^{2}\left(  \alpha\left(  2\lambda
^{2}-3\right)  -\lambda\right)  +\sigma_{1}\sigma_{2}\left(  3\lambda\left(
\lambda-2\alpha\left(  \lambda^{2}-3\right)  \right)  -\frac{9}{2}\right)
+\frac{9}{8}\sigma_{2}^{2}\left(  2\alpha\left(  2\lambda^{2}-9\right)
\lambda-2\lambda^{2}+3\right)  +\mathcal{O}(3),\\
&  \sigma_{2}^{\prime}=\sigma_{3}+\mathcal{O}(3),\\
&  \sigma_{3}^{\prime2}=\sigma_{1}\sigma_{3}(\alpha\lambda+1)-\frac{3}%
{2}\sigma_{2}\sigma_{3}\left(  4\lambda^{2}(\alpha\lambda+1)+3\right)
+\mathcal{O}(3)
\end{align}
We propose the following anzats:
\end{subequations}
\begin{equation}
\sigma_{1}=\frac{a_{1}}{\tau}+\frac{a_{2}}{\tau^{2}}+\mathcal{O}(\tau
^{-3}),\quad\sigma_{2}=\frac{b_{1}}{\tau}+\frac{b_{2}}{\tau^{2}}%
+\mathcal{O}(\tau^{-3}),\quad\sigma_{3}=\frac{c_{1}}{\tau}+\frac{c_{2}}%
{\tau^{2}}+\mathcal{O}(\tau^{-3})
\end{equation}
as $\tau\rightarrow\infty$. Substituting back and comparing the coefficient of
equal powers of $\tau$ we find the solutions:
\begin{subequations}
\begin{align}
&  a_{1}=-\frac{\sqrt{3b_{1}\left(  8\lambda\left(  -2\alpha\lambda
^{2}+6\alpha+\lambda\right)  +9b_{1}(2\lambda(\alpha(2\lambda(3\alpha
+\lambda)-9)-\lambda)+3)-12\right)  +4}-3b_{1}\left(  4\alpha\lambda
^{3}-12\alpha\lambda-2\lambda^{2}+3\right)  +2}{4\lambda\left(  \alpha\left(
2\lambda^{2}-3\right)  -\lambda\right)  },\nonumber\\
&  c_{1}=0,c_{2}=-b_{1},
\end{align}
or
\begin{align}
&  a_{1}=-\frac{-\sqrt{3b_{1}\left(  8\lambda\left(  -2\alpha\lambda
^{2}+6\alpha+\lambda\right)  +9b_{1}(2\lambda(\alpha(2\lambda(3\alpha
+\lambda)-9)-\lambda)+3)-12\right)  +4}-3b_{1}\left(  4\alpha\lambda
^{3}-12\alpha\lambda-2\lambda^{2}+3\right)  +2}{4\lambda\left(  \alpha\left(
2\lambda^{2}-3\right)  -\lambda\right)  },\nonumber\\
&  c_{1}=0,c_{2}=-b_{1}.
\end{align}
This approximated solution has the correct number of arbitrary constants:
$a_{2},b_{1},b_{2}$. Then, we obtain
\end{subequations}
\begin{subequations}
\begin{align}
&  x=\frac{(3b_{1}-2a_{1})\lambda}{\sqrt{6}\tau}+\frac{(3b_{2}-2a_{2})\lambda
}{\sqrt{6}\tau^{2}}+\mathcal{O}(\tau^{-3}),\\
&  y=\frac{4(2a_{1}-3b_{1})^{2}\alpha\lambda^{3}+2(2a_{1}-3b_{1})^{2}%
\lambda^{2}+3b_{1}(9b_{1}-4)}{12\tau^{2}}+\mathcal{O}(\tau^{-3}),\\
&  z=\frac{3b_{1}}{2\tau}+\frac{3b_{2}}{2\tau^{2}}+\mathcal{O}(\tau^{-3}).
\end{align}
From here it is clear that the origin attracts the orbits as $\tau
\rightarrow\infty$.

On the other hand, we get%

\end{subequations}
\begin{subequations}
\begin{align}
&  \dot{\phi}(t(\tau))=\frac{(3b_{1}-2a_{1})\lambda}{\tau}+\frac
{(3b_{2}-2a_{2})\lambda}{\tau^{2}}+\mathcal{O}(\tau^{-3}),\\
&  H(t(\tau))=\frac{3b_{1}}{2\tau}+\frac{3b_{2}}{2\tau^{2}}+\mathcal{O}%
(\tau^{-3}),\\
&  \phi(t(\tau))=\frac{\ln\left(  \frac{4V_{0}}{4(2a_{1}-3b_{1})^{2}%
\alpha\lambda^{3}+2(2a_{1}-3b_{1})^{2}\lambda^{2}+3b_{1}(9b_{1}-4)}\right)
+2\ln\tau}{\lambda}+\mathcal{O}(\tau^{-1}).
\end{align}

Furthermore
\end{subequations}
\begin{equation}
\frac{dt}{d\tau}=\frac{1}{\sqrt{H^{2}+1}}=1-\frac{3b_{1}}{4\tau}%
+\frac{3\left(  9b_{1}^{2}-8b_{2}\right)  }{32\tau^{2}}+\mathcal{O}(\tau^{-3})
\end{equation}
which implies
\begin{equation}
\Delta t=t(\tau)-t_{1}=\tau-\frac{3}{4}b_{1}\ln\tau+\frac{3\left(
8b_{2}-9b_{1}^{2}\right)  }{32\tau}+\mathcal{O}(\tau^{-2}).
\end{equation}
Inverting the above expression we have
\begin{equation}
\tau=\frac{3\left(  9b_{1}^{2}-8b_{2}\right)  }{32\Delta t}+\frac{9b_{1}%
^{2}\ln\Delta t}{16\Delta t}+\frac{3}{4}b_{1}\ln\Delta t +\Delta
t+\mathcal{O}(\Delta t^{-2}).
\end{equation}
Finally,
\begin{subequations}
\label{EqB15}%
\begin{align}
&  \dot{\phi}(t)=-\frac{3b_{1}\lambda(3b_{1}-2a_{1})\ln t}{4t^{2}}%
+\frac{\lambda(3b_{1}-2a_{1})}{t}+\frac{\lambda(3b_{2}-2a_{2})}{t^{2}%
}+\mathcal{O}(t^{-3}),\\
&  H(t)=-\frac{9b_{1}^{2}\ln t}{8t^{2}}+\frac{3b_{1}}{2t}+\frac{3b_{2}}%
{2t^{2}}+\mathcal{O}(t^{-3}),\\
&  \phi(t)=\frac{\ln\left( \frac{4 V_{0}}{4 \alpha\lambda^{3} (2 a_{1}-3
b_{1})^{2}+2 \lambda^{2} (2 a_{1}-3 b_{1})^{2}+3b_{1} (9 b_{1}-4)}\right)
}{\lambda}+\frac{2 \ln t}{\lambda}+\mathcal{O}(t^{-1}),\\
&  a(t)=a_{0} t^{\frac{3 b_{1}}{2}} \left( 1+\frac{3 \left( 3 {b_{1}}^{2}\ln t
+3 {b_{1}}^{2}-4 {b_{2}}\right) }{8 t}+\mathcal{O}\left( t^{-2}\right) \right)
.
\end{align}

For simplicity we set $t_{1}=0,\Delta t=t$.

On the other hand, the new conservation law
\end{subequations}
\[
\frac{g_{0} a^{3} e^{\lambda\phi} {\dot\phi}^{2} \left( 6 H-\lambda\dot
\phi\right) }{\lambda}+\frac{2 a^{3} \left( \lambda H-\dot\phi\right)
}{\lambda}+I_{1}=0,
\]
has to be satisfied. Substituting the expression \eqref{EqB15}, we obtain
\begin{align*}
& \frac{I_{1}}{a_{0}^{3}}+\frac{1}{8} t^{\frac{9 b_{1}}{2}-2} \left( 9 (4
a_{1}-3 b_{1}) \left( 3 b_{1}^{2}-4 b_{2}\right) +3 b_{1} (9 b_{1}-2) (4
a_{1}-3 b_{1}) \ln t +32 a_{2}-24 b_{2}\right) \\ &  +(4 a
_{1}-3 b_{1}) t^{\frac{9 b_{1}}{2}-1}+\mathcal{O}\left( t^{-3}\right) =0.
\end{align*}
where $a_{1}=a_{1}(b_{1}, \lambda, \alpha)$. Thus, for eliminating the term
$\propto t^{\frac{9 {b_{1}}}{2}-1}$, we have the following choices:

\begin{itemize}
\item We set the exponent to zero, i.e., we set $b_{1}=\frac{2}{9}$. Then
$a_{1}$ is reduced to the two values:
\[
a_{1}^{\pm}(\lambda,\alpha)= -\frac{2 \lambda\left( \lambda-2 \alpha\left(
\lambda^{2}-3\right) \right) \pm\sqrt{6} \sqrt{\lambda\left( \alpha\left( 6
\alpha\lambda-2 \lambda^{2}+3\right) +\lambda\right) }}{6 \lambda\left(
\alpha\left( 2 \lambda^{2}-3\right) -\lambda\right) }.
\]
Substituting $b_{1}=\frac{2}{9}$, and specifying $a_{1}=a_{1}^{\pm}(\lambda,
\alpha)$, the constraint becomes
\[
\frac{I_{1}}{{a_{0}}^{3}}+\frac{a_{1} \left( \frac{2}{3}-18 b_{2}\right) +4
a_{2}-\frac{1}{9}}{t}+4 a_{1}-\frac{2}{3}+\mathcal{O}\left( t^{-3}\right) =0.
\]
Now, in order to satisfy the above expressions up to the order $\mathcal{O}%
\left( t^{-3}\right) $, we impose the conditions
\[
{a_{1}}  (6-162 {b_{2}})+36 {a_{2}}-1=0, \quad4 {a_{1}}-\frac{2 }{3}%
+\frac{I_{1}}{{a_{0}}^{3}}=0,
\]
that has four arbitrary constants $a_{2}, b_{2}, a_{0}, I_{1}$, from which we
cannot eliminate simultaneously $a_{2}$ and $b_{2}$, or $I_{1}$ and $a_{0}$.
This leads to a two-parameter family of solutions. We solve for $a_{2}$ and
$I_{1}$ to keep the parameters $b_{2}$ and $a_{0}$. Finally,
\begin{subequations}
\begin{align}
& \dot\phi(t)=\frac{\left( \frac{2}{3}-2 a_{1}\right)  \lambda}{t}%
+\frac{\lambda\left( -18 a_{2}+27 b_{2}-(1-3 a_{1}) \ln t\right) }{9t^{2}%
}+\mathcal{O}\left( t^{-3}\right) ,\\
& H(t)= \frac{1}{3 t}+\frac{27 b_{2}-\ln t}{18 t^{2}}+\mathcal{O}\left(
t^{-3}\right) ,\\
& \phi(t)=\frac{\ln\left( \frac{9 V_{0}}{2 (1-3 a_{1})^{2} \lambda^{2} (2
\alpha\lambda+1)-3}\right) +2 \ln t}{\lambda}+\mathcal{O}\left( t^{-1}\right)
,\\
& a(t)=a_{0} \sqrt[3]{t}-\frac{1}{18} \left( a_{0} \left( 27 b_{2}-\ln
t-1\right) \right)  t^{-\frac{2}{3}}+\mathcal{O}\left( t^{-\frac{5}{3}}\right)
.
\end{align}

\item We can equate the coefficient of $t^{\frac{9 {b_{1}}}{2}-1}$ to zero,
that is, to solve $(4 a_{1}-3 b_{1})=0$ for $b_{1}$. The next term on the
expansion will be $(4 a_{2}-3 b_{2}) t^{\frac{9 b_{1}}{2}-2}$. Setting $(4
a_{2}-3 b_{2})=0$, we obtain the constraints $I_{1}=0$ and $b_{1}=\frac{4}{3}
a_{1}, b_{2}=\frac{4}{3}a_{2}$, $a_{1} \left( a_{1} \lambda\left( \alpha\left(
2 \lambda^{2}-15\right) -\lambda\right) +1\right) =0$.

\item Finally, we have the fine tuned solution $a_{1}=\frac{1}{3}$ and
$b_{1}=\frac{4}{9}$, that leads to the constraints $2 \alpha\lambda^{3}-15
\alpha\lambda-\lambda^{2}=0$ and $\frac{I_{1}}{{a_{0}}^{3}}+4 a_{2}-3 b_{2}=0$.
\end{subequations}
\end{itemize}

This point has not been obtained previously in \cite{genlyGL} and in
\cite{genlyGL2}, since in these works the authors used $H$-normalization,
which fails obviously when $H=0$.

For examining the stability of the fixed point $P_{8}$, we introduce the new
variable $\tilde{z}=z+1$. Then, the evolution equations \eqref{eq.38} can be
written as
\begin{equation}
x^{\prime}=x\left(  \frac{3}{2}-6\alpha\lambda\right)  +\sqrt{\frac{3}{2}%
}\lambda y+\mathcal{O}(2),\quad y^{\prime}=-3y+\mathcal{O}(2),\quad\tilde
{z}^{\prime}=-3\tilde{z}+\mathcal{O}(2).
\end{equation}
For examining the stability of the the fixed point $P_{9}$, we introduce the
new variable $\tilde{z}=z-1$. Then, the evolution equations \eqref{eq.38} can
be written as
\begin{equation}
x^{\prime}=-x\left(  \frac{3}{2}-6\alpha\lambda\right)  +\sqrt{\frac{3}{2}%
}\lambda y+\mathcal{O}(2),\quad y^{\prime}=3y+\mathcal{O}(2),\quad\tilde
{z}^{\prime}=3\tilde{z}+\mathcal{O}(2).
\end{equation}
From this linearized equations we extract the eigenvalues for $P_{8}$ and
$P_{9}$.

\section{Center Manifold for $P_{2}$}

\label{AppP2} The point $P_{3}$ has the opposite stability behavior.

For investigating the stability of the center manifold of $P_{2}$ we introduce
the new variables\newline$w=y, \quad u=z+1, \quad v= x-\frac{\sqrt{6}}%
{\lambda} z.$ Hence, we have the evolution equations
\begin{subequations}
\label{eqA2}%
\begin{align}
&  w^{\prime}= 12 u w+\lambda\left(  \sqrt{6} v w-\frac{w^{2}}{2 \alpha
}\right)  +\frac{\lambda^{3} w^{2}}{12 \alpha} +\mathcal{O}(3),\\
&  u^{\prime2}+2 \sqrt{6} \lambda u v +\mathcal{O}(3),\\
&  v^{\prime}=-3 v+9 u v+\frac{\lambda^{2} w (u+w)}{4 \sqrt{6} \alpha}%
-\frac{\sqrt{\frac{3}{2}} u w}{2 \alpha}+\frac{\lambda\left(  72 \sqrt{6}
\alpha^{2} v^{2}-24 \alpha v w+\sqrt{6} w^{2}\right)  }{48 \alpha^{2}}
+\nonumber\\
&  -\frac{\lambda^{3} w \left(  \sqrt{6} w-6 \alpha v\right)  }{72 \alpha^{2}%
}+\frac{\lambda^{5} w^{2}}{96 \sqrt{6} \alpha^{2}} +\mathcal{O}(3).
\end{align}
The center manifold of $P_{2}$ is then given locally by\newline$\{(w,u,v):
u=h_{1}(w), \quad v=h_{2}(w), \quad h_{1}(0)=0, \quad h_{2}(0)=0, \quad
h_{1}^{\prime}(0)=0, \quad h^{\prime}_{2}(0)=0, \quad|w|<\delta\}$, where
$\delta$ is small enough, and the functions $h_{1}, h_{2}$ satisfy the
equations
\end{subequations}
\begin{subequations}
\label{eqA4}%
\begin{align}
&  h_{1}^{\prime}(w) \left(  -12 w h_{1}(w)-\sqrt{6} \lambda w h_{2}%
(w)-\frac{\lambda\left(  \lambda^{2}-6\right)  w^{2}}{12 \alpha}\right)
+h_{1}(w) \left(  2 \sqrt{6} \lambda h_{2}(w)-6\right)  +15 h_{1}(w)^{2}=0,\\
&  h_{2}^{\prime}(w) \left(  -12 w h_{1}(w)-\sqrt{6} \lambda w h_{2}%
(w)-\frac{\lambda\left(  \lambda^{2}-6\right)  w^{2}}{12 \alpha}\right)
+h_{1}(w) \left(  9 h_{2}(w)+\frac{\left(  \lambda^{2}-6\right)  w}{4 \sqrt{6}
\alpha}\right) \nonumber\\
&  +h_{2}(w) \left(  \frac{\lambda\left(  \lambda^{2}-6\right)  w}{12 \alpha
}-3\right)  +3 \sqrt{\frac{3}{2}} \lambda h_{2}(w)^{2}+\frac{\lambda w^{2}
\left(  24 \alpha\lambda+\lambda^{4}-8 \lambda^{2}+12\right)  }{96 \sqrt{6}
\alpha^{2}}=0.
\end{align}

We have to solve this equations using Taylor series up to third order since
the equations \eqref{eqA2} were truncated at third order. Assuming $h_{1}(w)=a
w^{2}+\mathcal{O}(w)^{3}$ and $h_{2}(w)=b w^{2}+\mathcal{O}(w)^{3}$ (we start
at second order in $w$ since the conditions $h_{1}(0)=0, \quad h_{2}(0)=0,
\quad h_{1}^{\prime}(0)=0, \quad h^{\prime}_{2}(0)=0$ has to be satisfied),
substituting back into \eqref{eqA4}, and equating the coefficients with the
same powers of $w$ we find $a=0, b= \frac{\lambda\left(  24 \alpha
\lambda+\lambda^{4}-8 \lambda^{2}+12\right)  }{288 \sqrt{6} \alpha^{2}}$. The
evolution on the center manifold is governed by $w^{\prime}=\frac
{\lambda\left(  \lambda^{2}-6\right)  w^{2}}{12 \alpha}+\frac{\lambda^{2}
w^{3} \left(  24 \alpha\lambda+\lambda^{4}-8 \lambda^{2}+12\right)  }{288
\alpha^{2}}+O\left(  w^{4}\right) ,$ a ``gradient'' equation with potential
$W(w)=-\frac{\lambda^{2} w^{4} \left(  24 \alpha\lambda+\lambda^{4}-8
\lambda^{2}+12\right)  }{1152 \alpha^{2}}-\frac{\lambda\left(  \lambda
^{2}-6\right)  w^{3}}{36 \alpha}$.
Since the first non-zero derivative at $w=0$ is the third one, then the point
$w=0$ is an inflection point of $W(u)$. Hence, for $\frac{\lambda\left(
\lambda^{2}-6\right)  }{12 \alpha}>0$, the solutions with $w(0) > 0$ leave the
origin (and go off to $\infty$ in finite time) whereas the solutions with
$w(0) < 0$ approach the equilibrium as time passes. For $\frac{\lambda\left(
\lambda^{2}-6\right)  }{12 \alpha}<0$, the solutions with $w(0) > 0$ approach
the equilibrium as time passes whereas the solutions with $w(0) < 0$ leave the
origin (and go off to $\infty$ in finite time). Such an equilibrium with
one-sided stability is sometimes said to be semi-stable. However, since we
have to restrict our attention to the region $w\geq0$ (since $y\geq0$), we
have that $P_{2}$ is a saddle for $\frac{\lambda\left(  \lambda^{2}-6\right)
}{12 \alpha}>0$, and stable for $\frac{\lambda\left(  \lambda^{2}-6\right)
}{12 \alpha}<0$. But $\alpha$ is always non-negative, thus, $P_{2}$ is stable
for $\lambda<-\sqrt{6}$ or $0<\lambda<\sqrt{6}$, and saddle for $-\sqrt
{6}<\lambda<0$ or $\lambda>\sqrt{6}$.

\section{Center Manifold for $P_{4}$}

\label{AppP4} We discuss in more detail the stability of the fixed point
$P_{4}$. The point $P_{5}$ has the opposite stability behavior.

For investigating the stability of the center manifold of $P_{4}$ we introduce
the new variables\newline$w=z,\quad u=y-2z+1,\quad v=x.$ Hence, we have the
evolution equations
\end{subequations}
\begin{subequations}
\label{eqA6}%
\begin{align}
&  w^{\prime}=-3uw+\mathcal{O}(3),\\
&  u^{\prime}=-3\left(  u^{2}+uw+u-v^{2}-w^{2}\right)  +\mathcal{O}(3),\\
&  v^{\prime}=-\frac{3}{2}v\left(  u+2\left(  \sqrt{6}\alpha v+w+1\right)
\right)  +\mathcal{O}(3).
\end{align}
The center manifold of $P_{4}$ is then given locally by\newline%
$\{(w,u,v):u=h_{1}(w),\quad v=h_{2}(w),\quad h_{1}(0)=0,\quad h_{2}(0)=0,\quad
h_{1}^{\prime}(0)=0,\quad h_{2}^{\prime}(0)=0,\quad|w|<\delta\}$, where
$\delta$ is small enough, and the functions $h_{1},h_{2}$ satisfy the
equations
\end{subequations}
\begin{subequations}
\label{eqA8}%
\begin{align}
&  3wh_{1}(w)h_{1}^{\prime}(w)-3h_{1}(w)^{2}-3(w+1)h_{1}(w)+3h_{2}%
(w)^{2}+3w^{2}=0,\\
&  3wh_{1}(w)h_{2}^{\prime}(w)-\frac{3}{2}h_{1}(w)h_{2}(w)-3\sqrt{6}\alpha
h_{2}(w)^{2}-3(w+1)h_{2}(w)=0.
\end{align}

We have to solve this equations using Taylor series up to third order since
the equations \eqref{eqA6} were truncated at third order. Assuming
$h_{1}(w)=aw^{2}+\mathcal{O}(w)^{3}$ and $h_{2}(w)=bw^{2}+\mathcal{O}(w)^{3}$
(we start at second order in $w$ since the conditions $h_{1}(0)=0,\quad
h_{2}(0)=0,\quad h_{1}^{\prime}(0)=0,\quad h_{2}^{\prime}(0)=0$ has to be
satisfied), substituting back into \eqref{eqA6}, and equating the coefficients
with the same powers of $w$ we find $a=1,b=0$. The evolution on the center
manifold is governed by $w^{\prime}=-3w^{3}.$ The equilibrium $w=0$ is then
asymptotically stable. Furthermore, perturbations from the equilibrium grow or
decay algebraically in time, not exponentially as in the usual linear
stability analysis. This is the same result obtained previously in
\cite{genlyGL} and in \cite{genlyGL2}, for the de-Sitter (constant potential) solution.

\section{Center manifold for the line of fixed points $P_{14}(z_{c})$}

\label{AppP14} The line of fixed points $P_{14}(z_{c}):\left(  0,z_{c}%
^{2},z_{c}\right)  $ exists for $\lambda=0$. The eigenvalues are $\left(
0,-3z_{c},-3z_{c}\right)  $. Thus, it is nonhyperbolic. For investigating the
stability of the center manifold of $P_{14}(z_{c})$ we introduce the new
variables $w=\frac{-(y+1) {z_{c}}^{2}+y+2 z {z_{c}}^{3}-{z_{c}}^{4}}{2 {z_{c}%
}}, u=\frac{\left( {z_{c}}^{2}-1\right)  (y+{z_{c}} ({z_{c}}-2 z))}{2 {z_{c}}%
}, v=x.$ Hence, we have the evolution equations
\end{subequations}
\begin{subequations}
\label{eqE6}%
\begin{align}
&  w^{\prime}=3 u^{2} {z_{c}}^{2}+\frac{3 u w \left( {z_{c}}^{2}-3\right)
{z_{c}}^{2}}{{z_{c}}^{2}-1}+\mathcal{O}(3),\\
&  u^{\prime}=-\frac{3 u^{2} \left( {z_{c}}^{4}+4 {z_{c}}^{2}-1\right) }{2
\left( {z_{c}}^{2}-1\right) }+u (-3 w-3 {z_{c}})+\frac{3}{2} v^{2} \left(
{z_{c}}^{2}-1\right) +\frac{3}{2} w^{2} \left( {z_{c}}^{2}-1\right)
+\mathcal{O}(3),\\
&  v^{\prime}=\frac{u v \left( 3-6 {z_{c}}^{2}\right) }{{z_{c}}^{2}-1}-3
\sqrt{6} \alpha v^{2}+v (-3 w-3 {z_{c}}) +\mathcal{O}(3).
\end{align}
The center manifold of $P_{14}$ is then given locally by\newline%
$\{(w,u,v):u=h_{1}(w),\quad v=h_{2}(w),\quad h_{1}(0)=0,\quad h_{2}(0)=0,\quad
h_{1}^{\prime}(0)=0,\quad h_{2}^{\prime}(0)=0,\quad|w|<\delta\}$, where
$\delta$ is small enough, and the functions $h_{1},h_{2}$ satisfy the
equations
\end{subequations}
\begin{subequations}
\label{eqE8a}%
\begin{align}
\left( -3 {z_{c}}^{2} h_{1}(w)^{2}-\frac{3 w \left( {z_{c}}^{2}-3\right)
{z_{c}}^{2} h_{1}(w)}{{z_{c}}^{2}-1}\right)  h_{1}^{\prime}(w)-\frac{3 \left(
{z_{c}}^{4}+4 {z_{c}}^{2}-1\right)  h_{1}(w)^{2}}{2 \left( {z_{c}}%
^{2}-1\right) }-3 h_{1}(w) (w+{z_{c}})  & \nonumber\\
+\frac{3}{2} \left( {z_{c}}^{2}-1\right)  h_{2}(w)^{2}+\frac{3}{2} w^{2}
\left( {z_{c}}^{2}-1\right)   &  =0,
\end{align}
\begin{align}
\label{eqE8aa}\left( -3 {z_{c}}^{2} h_{1}(w)^{2}-\frac{3 w \left( {z_{c}}%
^{2}-3\right)  {z_{c}}^{2} h_{1}(w)}{{z_{c}}^{2}-1}\right)  h_{2}^{\prime
}(w)+\frac{\left( 3-6 {z_{c}}^{2}\right)  h_{1}(w) h_{2}(w)}{{z_{c}}^{2}-1} +
& \nonumber\\
-3 \sqrt{6} \alpha h_{2}(w)^{2}-3 h_{2}(w) (w+{z_{c}})  &  =0.
\end{align}

We have to solve this equations using Taylor series up to third order since
the equations \eqref{eqE6} were truncated at third order. Assuming
$h_{1}(w)=aw^{2}+\mathcal{O}(w)^{3}$ and $h_{2}(w)=bw^{2}+\mathcal{O}(w)^{3}$
(we start at second order in $w$ since the conditions $h_{1}(0)=0,\quad
h_{2}(0)=0,\quad h_{1}^{\prime}(0)=0,\quad h_{2}^{\prime}(0)=0$ has to be
satisfied), substituting back into \eqref{eqE6}, and equating the coefficients
with the same powers of $w$ we find $a=\frac{{z_{c}}^{2}-1}{2 {z_{c}}},b=0$.
The evolution on the center manifold is governed by $w^{\prime}=-\frac{3}{2}
w^{3} {z_{c}} \left( 3-{z_{c}}^{2}\right) w^{3}.$ The equilibrium $w=0$ is
then asymptotically stable for $0<{z_{c}}\leq1$ and asymptotically unstable
for $-1\leq{z_{c}}<0$. As before, perturbations from the equilibrium grow or
decay algebraically in time.

\section{Center manifold for the lines of fixed points $P_{15,16}(z_{c})$}

\label{AppP15-P16}

The lines of fixed points $P_{15,16}(z_{c}):\left( \beta{z_{c}},\sqrt{6}%
\alpha\beta{z_{c}}^{2},z_{c}\right)  $ exists for $\lambda=0,\beta=\beta
_{1,2}$, where $\beta_{1,2}=\frac{1}{\sqrt{2}}\left(  \sqrt{3}\alpha\mp
\sqrt{3\alpha^{2}-2}\right) $, respectively. The eigenvalues are $\left(
0,-3z_{c},-3z_{c}\right)  $. Thus, they are nonhyperbolic. The two cases can
be treated as one case under the choice $\alpha= \frac{\beta^{2}+1}{\sqrt{6}
\beta}$ (this is equivalent to setting $\beta=\frac{1}{\sqrt{2}}\left(
\sqrt{3}\alpha\mp\sqrt{3\alpha^{2}-2}\right) $).

For investigating the stability of the center manifold of $P_{15,16}(z_{c})$
we introduce the new variables\newline$w=\frac{y-y {z_{c}}^{2}}{2 \beta^{2}
{z_{c}}+2 {z_{c}}}-\frac{1}{2} {z_{c}} \left( -2 z {z_{c}}+{z_{c}}%
^{2}+1\right) , u=\frac{\left( {z_{c}}^{2}-1\right)  \left( y-\left( \beta
^{2}+1\right)  {z_{c}} (2 z-{z_{c}})\right) }{2 \left( \beta^{2}+1\right)
{z_{c}}}, v=x+\frac{\beta\left( \frac{y \left( {z_{c}}^{2}-1\right) }%
{\beta^{2}+1}+{z_{c}}^{2} \left( -2 z {z_{c}}+{z_{c}}^{2}-1\right) \right) }{2
{z_{c}}}$. Hence, we have the evolution equations
\end{subequations}
\begin{subequations}
\label{eqF6}%
\begin{align}
&  w^{\prime}=3 u^{2} {z_{c}}^{2}+\frac{3 u w \left( {z_{c}}^{2}-3\right)
{z_{c}}^{2}}{{z_{c}}^{2}-1}+\mathcal{O}(3),\\
&  u^{\prime}=\frac{3 u^{2} \left( 2 \beta^{2}+{z_{c}}^{4}+\left( 4-6
\beta^{2}\right)  {z_{c}}^{2}-1\right) }{2 \left( \beta^{2}-1\right)  \left(
{z_{c}}^{2}-1\right) }+u \left( -\frac{3 \beta v \left( {z_{c}}^{2}-1\right)
}{\beta^{2}-1}-3 w-3 {z_{c}}\right) +\frac{3 v^{2} \left( {z_{c}}^{2}-1\right)
}{2 \left( \beta^{2}-1\right) }\nonumber\\
&  +\frac{3}{2} w^{2} \left( {z_{c}}^{2}-1\right) +\mathcal{O}(3),\\
&  v^{\prime}=\frac{3 \beta u^{2} \left( {z_{c}}^{2}+1\right)  \left( {z_{c}%
}^{4}+2 \left( \beta^{2}-2\right)  {z_{c}}^{2}+1\right) }{2 \left( \beta
^{2}-1\right)  \left( {z_{c}}^{2}-1\right) ^{2}}+\frac{3 u v \left( {z_{c}%
}^{2} \left( 4-\beta^{2} \left( {z_{c}}^{2}+4\right) \right) +1\right)
}{\left( \beta^{2}-1\right)  \left( {z_{c}}^{2}-1\right) }+\frac{v^{2} \left(
3 \beta^{2} \left( {z_{c}}^{2}+3\right) -6\right) }{2 \beta\left( \beta
^{2}-1\right) }\nonumber\\
&  +v (-3 w-3 {z_{c}}) +\frac{3}{2} \beta w^{2} \left( {z_{c}}^{2}-1\right)
+\mathcal{O}(3).
\end{align}
The center manifold of $P_{14}$ is then given locally by \newline%
$\{(w,u,v):u=h_{1}(w),\quad v=h_{2}(w),\quad h_{1}(0)=0,\quad h_{2}(0)=0,\quad
h_{1}^{\prime}(0)=0,\quad h_{2}^{\prime}(0)=0,\quad|w|<\delta\}$, where
$\delta$ is small enough, and the functions $h_{1},h_{2}$ satisfy the
equations
\end{subequations}
\begin{subequations}
\label{eqF8a}%
\begin{align}
& \left( -3 {z_{c}}^{2} {h_{1}}(w)^{2}-\frac{3 w \left( {z_{c}}^{2}-3\right)
{z_{c}}^{2} {h_{1}}(w)}{{z_{c}}^{2}-1}\right)  {h_{1}}^{\prime}(w)+{h_{1}}(w)
\left( -\frac{3 \beta\left( {z_{c}}^{2}-1\right)  {h_{2}}(w)}{\beta^{2}-1}-3
(w+{z_{c}})\right) \nonumber\\
&  +\frac{3 {h_{1}}(w)^{2} \left( 2 \beta^{2}+{z_{c}}^{4}+\left( 4-6 \beta
^{2}\right)  {z_{c}}^{2}-1\right) }{2 \left( \beta^{2}-1\right)  \left(
{z_{c}}^{2}-1\right) }+\frac{3 \left( {z_{c}}^{2}-1\right)  {h_{2}}(w)^{2}}{2
\left( \beta^{2}-1\right) }+\frac{3}{2} w^{2} \left( {z_{c}}^{2}-1\right)  =0,
\end{align}
\begin{align}
\label{eqF8aa} & \left( -3 {z_{c}}^{2} {h_{1}}(w)^{2}-\frac{3 w \left( {z_{c}%
}^{2}-3\right)  {z_{c}}^{2} {h_{1}}(w)}{{z_{c}}^{2}-1}\right)  {h_{2}}%
^{\prime}(w)+\frac{3 {h_{1}}(w) {h_{2}}(w) \left( {z_{c}}^{2} \left(
4-\beta^{2} \left( {z_{c}}^{2}+4\right) \right) +1\right) }{\left( \beta
^{2}-1\right)  \left( {z_{c}}^{2}-1\right) }\nonumber\\
&  +\frac{3 \beta\left( {z_{c}}^{2}+1\right)  {h_{1}}(w)^{2} \left( {z_{c}%
}^{4}+2 \left( \beta^{2}-2\right)  {z_{c}}^{2}+1\right) }{2 \left( \beta
^{2}-1\right)  \left( {z_{c}}^{2}-1\right) ^{2}}+\frac{3 {h_{2}}(w)^{2} \left(
\beta^{2} \left( {z_{c}}^{2}+3\right) -2\right) }{2 \beta\left( \beta
^{2}-1\right) }-3 {h_{2}}(w) (w+{z_{c}})\nonumber\\
&  +\frac{3}{2} \beta w^{2} \left( {z_{c}}^{2}-1\right)  =0.
\end{align}

We have to solve this equations using Taylor series up to third order since
the equations \eqref{eqF6} were truncated at third order. Assuming
$h_{1}(w)=aw^{2}+\mathcal{O}(w)^{3}$ and $h_{2}(w)=bw^{2}+\mathcal{O}(w)^{3}$
(we start at second order in $w$ since the conditions $h_{1}(0)=0,\quad
h_{2}(0)=0,\quad h_{1}^{\prime}(0)=0,\quad h_{2}^{\prime}(0)=0$ has to be
satisfied), substituting back into \eqref{eqF6}, and equating the coefficients
with the same powers of $w$ we find $a=\frac{{z_{c}}^{2}-1}{2 {z_{c}}}, b=
\frac{\beta( {z_{c}}^{2}-1) }{2 {z_{c}}}$. The evolution on the center
manifold is governed by $w^{\prime}=-\frac{3}{2} w^{3} {z_{c}} \left(
3-{z_{c}}^{2}\right) w^{3}.$ The equilibrium $w=0$ is then asymptotically
stable for $0<{z_{c}}\leq1$ and asymptotically unstable for $-1\leq{z_{c}}<0$.
As before, perturbations from the equilibrium grow or decay algebraically in time.


\end{subequations}

\end{document}